\pdfoutput=1

\documentclass[journal]{IEEEtran}

\usepackage{Preamble}

\newcommand{\LF}{LF\xspace}

\newcommand{\DL}[1][]{DL#1\xspace}
\newcommand{\DG}[1][]{DG#1\xspace}

\newcommand{\ARIMA}{ARIMA\xspace}

\newcommand{\WLS}[1][]{WLS#1\xspace}		

\newcommand{\SE}[1][]{SE#1\xspace}				
\newcommand{\LSE}[1][]{LSE#1\xspace}			
\newcommand{\RTSE}[1][]{RTSE#1\xspace}		

\newcommand{\ADN}[1][]{ADN#1\xspace}

\newcommand{\DKF}[1][]{DKF#1\xspace}
\newcommand{\SDKF}[1][]{SDKF#1\xspace}

\newcommand{\TB}[1][]{TB\xspace}				
\newcommand{\MUT}{MUT\xspace}				
\newcommand{\GM}{GM\xspace}				

\newcommand{\SGL}{SGL\xspace}	
\newcommand{\DBL}{DBL\xspace}	

\newcommand{\EPFL}{EPFL\xspace}
\newcommand{\IEEE}{IEEE\xspace}
\newcommand{\NI}{NI\xspace}

\newcommand{\CRIO}[1][]{cRIO\xspace}

\newcommand{\CPU}[1][]{CPU#1\xspace}	
\newcommand{\FPGA}[1][]{FPGA#1\xspace}
\newcommand{\DSP}[1][]{DSP#1\xspace}		

\newcommand{\DMA}[1][]{DMA#1\xspace}	
\newcommand{\FIFO}[1][]{FIFO#1\xspace}	
\newcommand{\RAM}[1][]{RAM#1\xspace}	
\newcommand{\LUT}[1][]{LUT#1\xspace}		
\newcommand{\FF}[1][]{FF#1\xspace}			

\newcommand{\PMU}[1][]{PMU#1\xspace}	

\newcommand{\Radian}{\text{rad}}
\newcommand{\PerUnit}{\text{pu}}
\newcommand{\kV}{\text{kV}}

\newcommand{\MVA}{\text{MVA}}

\newcommand{\Buses}{\mathcal{B}}
\newcommand{\Phases}{\mathcal{P}}
\newcommand{\Instruments}{\mathcal{M}}

% ...

\newcommand{\ArgMin}{\operatorname{arg}\,\operatorname{min}}

\newcommand{\Tensor}[1]{\boldsymbol{\mathbf{#1}}}

\newcommand{\Expectation}[1]{\mathbb{E}\left[#1\right]}

\newcommand{\Normal}[1]{\mathcal{N}(#1)}

\newcommand{\Real}[1]{\Re\{#1\}}
\newcommand{\Imaginary}[1]{\Im\{#1\}}

\newcommand{\StateVariable}{x}
\newcommand{\MeasurementVariable}{z}

\newcommand{\EstimationErrorCovariance}{P}
\newcommand{\ProcessNoiseCovariance}{Q}
\newcommand{\MeasurementNoiseCovariance}{R}

\newcommand{\Estimate}[1]{\widehat{#1}}

\newcommand{\nx}{\Tensor{w}}
\newcommand{\nz}{\Tensor{v}}

\newcommand{\e}{\Tensor{e}}

\newcommand{\xa}{\Tensor{\StateVariable}}					
\newcommand{\xe}{\Estimate{\Tensor{\StateVariable}}}	
\newcommand{\ua}{\Tensor{u}}
\newcommand{\za}{\Tensor{\MeasurementVariable}}
\newcommand{\ze}{\widehat{\Tensor{z}}}

\newcommand{\dx}{d\Tensor{\StateVariable}}
\newcommand{\dz}{d\Tensor{\MeasurementVariable}}

\newcommand{\Se}{\Tensor{\EstimationErrorCovariance}}
\newcommand{\Sx}{\Tensor{\ProcessNoiseCovariance}}
\newcommand{\Sz}{\Tensor{\MeasurementNoiseCovariance}}

\newcommand{\W}{\Tensor{W}}

\newcommand{\dSx}{d\Tensor{\EstimationErrorCovariance}}
\newcommand{\dSz}{d\Tensor{\MeasurementNoiseCovariance}}
\newcommand{\dTz}{\Tensor{C}}			

\newcommand{\K}{\Tensor{K}}
\newcommand{\Tz}{\Tensor{H}}

\newcommand{\row}{\operatorname{row}}

\newcommand{\BigPsi}{%
	\mathop%
	{%
		\vphantom{\sum}%
		\mathchoice%
			{\vcenter{\hbox{\huge\textPsi}}}%
			{\vcenter{\hbox{\Large\textPsi}}}%
			{\mathrm{\Psi}}%
			{\mathrm{\Psi}}%
	}%
	\displaylimits%
	}

\DeclareMathOperator*{\Product}{\BigPsi}

\newcommand{\Order}[1]{\mathcal{O}(#1)}

\newcommand{\Appendix}[1]{Appendix~\ref{#1}}
\newcommand{\Equation}[1]{(\ref{#1})}
\newcommand{\Figure}[1]{Fig.~\ref{#1}}
\newcommand{\Section}[1]{Section~\ref{#1}}
\newcommand{\Table}[1]{Table~\ref{#1}}

{
\theoremstyle{plain}
\newtheorem{Theorem}{Theorem}
\newtheorem{Algorithm}{Algorithm}
\newtheorem{Lemma}{Lemma}
\newtheorem{Hypothesis}{Hypothesis}
}

{
\theoremstyle{definition}
\newtheorem*{Proof}{Proof}
}

\newcommand{\QEDtotal}{\hfill\IEEEQED}
\newcommand{\QEDpartial}{\hfill\IEEEQEDopen}

\newcolumntype{a}{>{$}r<{$}@{}>{${}}l<{$}} 

\newcolumntype{L}[1]{>{\raggedright\let\newline\\\arraybackslash\hspace{0pt}}m{#1}}
\newcolumntype{C}[1]{>{\centering\let\newline\\\arraybackslash\hspace{0pt}}m{#1}}
\newcolumntype{R}[1]{>{\raggedleft\let\newline\\\arraybackslash\hspace{0pt}}m{#1}}

\begin{document}


\title{
	Sequential Discrete Kalman Filter for Real-Time State Estimation in Power Distribution Systems:
	Theory and Implementation
}



\author{
	Andreas~Martin~Kettner,~\IEEEmembership{Member,~IEEE}
	and~Mario~Paolone,~\IEEEmembership{Senior~Member,~IEEE,}
	\thanks{%
		The authors are with the {\'E}cole Polytechnique F{\'e}d{\'e}rale de Lausanne in Lausanne, Switzerland %
		(E-mail: \{andreas.kettner$\,$,$\,$mario.paolone\}@epfl.ch).%
	}
}

%
%


\markboth{}{}
%




\maketitle



\begin{abstract}
	This paper demonstrates the feasibility of implementing Real-Time State Estimators (\RTSE[s]) for 
	Active Distribution Networks (\ADN[s]) in Field-Programmable Gate Arrays (\FPGA[s])
	by presenting an operational prototype.
	The prototype is based on a Linear State Estimator (\LSE) that uses synchrophasor measurements
	from Phasor Measurement Units (\PMU[s]).
	The underlying algorithm is the Sequential Discrete Kalman Filter (\SDKF), an equivalent formulation of
	the Discrete Kalman Filter (\DKF) for the case of uncorrelated measurement noise.
	In this regard, this work formally proves the equivalence of the \SDKF and the \DKF, 
	and highlights the suitability of the \SDKF for an \FPGA implementation by means of a computational complexity analysis.
	The developed prototype is validated using a case study adapted from the IEEE 34-node distribution test feeder.
\end{abstract}



\begin{IEEEkeywords}
	Active Distribution Network (\ADN),
	Real-Time State Estimator (\RTSE),
	Phasor Measurement Unit (\PMU),
	Sequential Discrete Kalman Filter (\SDKF),
	Field-Programmable Gate Array (\FPGA)
\end{IEEEkeywords}

%
\IEEEpeerreviewmaketitle

\section{Introduction}
\label{Section:Introduction}

%
%
%



\IEEEPARstart{R}{ecently}, accurate \emph{Phasor Measurement Units} (\PMU[s]) capable of streaming
synchrophasors at refresh rates of some tens of frames per second
\cite{Standard:IEEE:2011:PMU:Measurement,Standard:IEEE:2011:PMU:Transfer} have become available.
Such devices can be implemented in dedicated and inexpensive hardware like
\emph{Field-Programmable Gate Arrays} (\FPGA[s]) \cite{Journal:Romano:2014:PMU:IpDFT:Implementation}.
Therefore, they may potentially be employed on a massive scale in power distribution systems.
In rectangular coordinates, the relation between the nodal voltage phasors and the nodal current phasors
or the branch current phasors is linear \cite{Book:Milano:2016:Estimation},
which enables the use of \emph{Linear State Estimators} (\LSE[s]).
Lately, this prospect has stimulated further developments in the field of \emph{Real-Time State Estimators} (\RTSE[s]).
Namely, it has been demonstrated that \LSE[s] on the basis of the \emph{Discrete Kalman Filter} (\DKF)
may attain execution times in the subsecond range, while considerably outperforming
traditional \LSE[s] based on \emph{Weighted Least Squares} (\WLS) in terms of estimation accuracy
\cite{Journal:Sarri:2016:SE:Linear:Performance}.
However, the implementation relies on a powerful \emph{Central Processing Unit} (\CPU)
for performing computationally heavy operations.
Therefore, there is a gap between the instrumentation and the state estimation in \emph{Active Distribution Networks} (\ADN[s])
with respect to dedicated hardware implementations, which this work aims to bridge.
In this regard, it is proposed to use the \emph{Sequential Discrete Kalman Filter} (\SDKF), 
because it solely involves elementary linear algebra operations,
which are suitable for an implementation in dedicated hardware.
The contributions of this paper are twofold.
Firstly, it is proven that the formulations of the power system state estimation problem
using the \SDKF and the \DKF are formally equivalent.
Secondly, an \FPGA implementation of an \RTSE for power distribution systems based on the \SDKF is presented and validated.
To the best of the authors' knowledge, this hardware implementation is the first of its kind.
In that sense, the content of this paper can facilitate the development of 
automation systems for \ADN[s] that rely on \RTSE[s].



The remainder of this publication is organized as follows:
First, a survey of state-of-the-art methods for state estimation in power transmission and distribution systems,
with particular reference to the requirements of \ADN applications, is presented in \Section{Section:Literature}.
Then, the formulation of the state estimation problem and the derivation of the \SDKF from the \DKF
are discussed in \Section{Section:Algorithm}.
Moreover, it is explained why the \SDKF, in contrast to the \DKF, is particularly suitable for an \FPGA implementation.
The developed hardware protopype is discussed in \Section{Section:Implementation},
and the results of the numerical validation and the scalability analysis are presented in \Section{Section:Validation}.
Finally, the conclusions are drawn in \Section{Section:Conclusion}.

\section{Literature Review}
\label{Section:Literature}

\subsection{State Estimation in Power Transmission Systems}
\label{Section:Literature:Transmission}

In power transmission systems, operators have been using state estimators in their control centers
for several decades \cite{Journal:Wu:2005:Estimation:History}.
Ever since the early works that pioneered state estimation in this field
\cite
{%
	Journal:Schweppe:1970:Estimation:Model:Exact,%
	Journal:Schweppe:1970:Estimation:Model:Approximate,%
	Journal:Schweppe:1970:Estimation:Implementation%
},
most of the research has focused on methods based on \WLS
\cite
{%
	Book:Grainger:1994:System,%
	Book:Monticelli:1999:Estimation,%
	Book:Abur:2004:Estimation%
}.
These approaches are \emph{static} in the sense that they do not take into account the time derivative of the system state.
Namely, the estimated state is computed as a \emph{maximum likelihood} fit to the measurements
available at a given time-step \cite{Conference:Schweppe:1974:Estimation}.
In general, both the state vector and the measurement vector may consist of nodal and branch quantities (i.e. voltages, currents, and powers), 
expressed in rectangular or polar coordinates, which results in a nonlinear measurement model \cite{Journal:Monticelli:2000:Estimation},
and requires the use of iterative methods for solving the \WLS problem.
The complexity of the solver methods, and the sheer size of the system models, ultimately limit the refresh rate of the estimated state
(typical refresh rates are in the order of minutes).
To increase the state refresh rate, one may partition the system and formulate a \emph{Multi-Area State Estimation} problem,
which can be parallelized using hierarchical or decentralized schemes (e.g. \cite{Journal:Gomez:2011:Estimation:Multiarea}).
Further acceleration is achieved with \emph{High-Performance Computing} in massively parallel computational hardware
\cite{Conference:Liu:2012:Estimation:HPC:Architecture,Book:Khaitan:2013:System:HPC}.
Such implementations may use general-purpose hardware, like a cluster of desktop machines \cite{Journal:Korres:2013:Estimation:HPC:CPU},
or exploit special-purpose components, such as \emph{Graphics Processing Units} \cite{Conference:Karimipour:2013:HPC:GPU}.

\subsection{State Estimation in Power Distribution Systems}
\label{Section:Literature:Distribution}


In power distribution systems, operation problems have historically been solved in the planning stage,
so that little intervention is needed during operation.
Due to the widespread connection of decentralized generation, distributed energy storage systems, and flexible loads,
there is presently an evolution from passive distribution networks towards \ADN[s] \cite{Magazine:CIGRE:2011:System:Future}.
Since these changes lead to frequent violations of operational constraints (e.g. voltage limits and line ampacities),
there is a need for \emph{Distribution Management Systems}, which allow to meet various real-time operation objectives
\cite{Report:CIGRE:2011:System:ADN}.
In view of the typical dynamics of \ADN[s], such tools need to rely on \RTSE[s]
with high refresh rates (e.g. tens of frames per second), low overall latency (e.g. tens of milliseconds), and high accuracy.
Recently, the emerging availability of \PMU[s] capable of streaming accurate synchrophasors at high refresh rates
\cite{Standard:IEEE:2011:PMU:Measurement,Standard:IEEE:2011:PMU:Transfer},
has supported such developments \cite{Journal:Liu:PMU:Placement}.
	

In analogy to the well-established approaches known from power transmission systems,
several works have adopted static state estimators based on \WLS for power distribution systems.
In particular, it has been recognized that estimators based on Linear \WLS perform better in terms of computation time
than those based on Nonlinear \WLS, because the problem can be solved analytically rather than numerically.
This is demonstrated in \cite{Journal:Lu:1995:Estimation:Linear:WLS}, where an \LSE based on current measurements
is compared against traditional nonlinear estimators based on power measurements.
A conceptually similar \LSE, which uses measurements of nodal voltages, nodal currents, and branch currents, 
is proposed in \cite{Journal:Haughton:2013:Estimation:Linear:WLS}.
Yet another \LSE, based on an alternative model whose state variables are the branch currents rather than the nodal voltages,
is discussed in \cite{Journal:Kelley:1995:Estimation:Linear:WLS}.


Other works have addressed the problem that, as previously mentioned, an estimator based on the \WLS is inherently static,
because it entirely ignores the dynamics of the system.
Although an early work \cite{Journal:Debs:1970:Estimation:Linear:DKF} has explored \emph{dynamic} state estimation 
using the \DKF in combination with a quasi-static model of the dynamics, the idea has received little attention until lately
\cite{Book:Milano:2016:Estimation}.
Recently, \cite{Journal:Sarri:2016:SE:Linear:Performance} has performed a thorough performance analysis
of \LSE[s] based on \WLS and the \DKF in terms of estimation accuracy and execution speed.
In particular, it has been demonstrated that the \DKF is capable of outperforming the \WLS in terms of estimation accuracy,
if the process noise associated with the quasi-static model is properly assessed
\cite{Conference:Zanni:2014:SE:DKF:Covariance,Journal:Zanni:2016:SE:DKF:Covariance}.
The execution times obtained for a \CPU implementation run on a desktop machine indicate that real-time operation is feasible,
which has also been verified experimentally in an actual feeder \cite{Conference:Pignati:2015:SE:DKF:Demonstrator}.
However, the speed is also subject to significant variation over time, i.e. the behavior is not deterministic.


Since the use of \RTSE[s] in power distribution systems requires a deployment on a massive scale,
the apparent need for powerful \CPU[s] presents a non negligible hindrance.
Firstly, the cost of the required hardware (e.g. a workstation) would simply render the application noncompetitive.
Furthermore, one would struggle to ensure reliable operation ``in the field'',
unless expensive custom hardware (e.g. a weatherproof industrial computer) is used.
Conversely, using weaker (cheaper) \CPU[s] would slow down the execution speed and increase the problems with jitter.
To ensure both fast and deterministic execution speed at low cost, one must resort to a dedicated hardware implementation.
In this context, one should note that classical High-Performance Computing solutions,
like the ones used in power transmission systems, are not an option,
because they also suffer from the previously discussed problems.
However, \FPGA implementations are a possible solution, since they may achieve high performance,
while being inexpensive and rugged.
For instance, an \FPGA prototype of a \PMU for power distribution systems has recently been
developed \cite{Journal:Romano:2014:PMU:IpDFT:Implementation}.
This work aims to close the gap between instrumentation and state estimation in terms of dedicated hardware implementations
by presenting an operational \FPGA prototype of an \RTSE.
The said prototype is based on the \SDKF, an equivalent formulation of the \DKF for the case of uncorrelated measurement noise,
which (in contrast to the latter) is suitable for this type of dedicated hardware.

\section{Algorithm Formulation}
\label{Section:Algorithm}

This section focuses on the theoretical aspects of this paper.
First, the models used for the dynamical system and the measurement system
are developed in \Section{Section:Algorithm:Model}.
Then, the formulas describing the \DKF and the \SDKF are summarized
in \Section{Section:Algorithm:Batch} and \Section{Section:Algorithm:Sequential}, respectively.
After, the proof of equivalence for the \DKF and the \SDKF
is presented in \Section{Section:Algorithm:Equivalence}.
Finally, the computational complexity of the different filters is analyzed in \Section{Section:Algorithm:Complexity}
in view of the deployment of the \SDKF into an \FPGA.




\subsection{System Model}
\label{Section:Algorithm:Model}



Consider an electrical grid with buses $b\in\Buses=\{1,\ldots,N\}$ and phases $p\in\Phases=\{1,2,3\}$.
Let $V_{b,p,k}$ and $I_{b,p,k}$ denote the phasors of the nodal voltage and nodal current
in phase $p\in\Phases$ of bus $b\in\Buses$.
Define $\Tensor{V}_{b,k}$ and $\Tensor{I}_{b,k}$ as the vectors of all nodal voltage and nodal current phasors
in bus $b\in\Buses$
\begin{equation}
	\Tensor{V}_{b,k}
	=	\left[
		\begin{array}{c}
			V_{b,1,k}\\
			V_{b,2,k}\\
			V_{b,3,k}
		\end{array}
		\right]
	~,~
	\Tensor{I}_{b,k}
	=	\left[
		\begin{array}{c}
			I_{b,1,k}\\
			I_{b,2,k}\\
			I_{b,3,k}
		\end{array}
		\right]
\end{equation}
Accordingly, the vectors $\Tensor{V}_{k}$ and $\Tensor{I}_{k}$ for the entire network are
\begin{equation}
	\Tensor{V}_{k}
	=	\left[
		\begin{array}{c}
			\Tensor{V}_{1,k}	\\
			\vdots				\\
			\Tensor{V}_{N,k}
		\end{array}
		\right]
	~,~
	\Tensor{I}_{k}
	=	\left[
		\begin{array}{c}
			\Tensor{I}_{1,k}	\\
			\vdots				\\
			\Tensor{I}_{N,k}
		\end{array}
		\right]
\end{equation}
Note that the vectors $\Tensor{V}_{k}$ and $\Tensor{I}_{k}$ are related as follows
\begin{equation}
	\Tensor{I}_{k} = \Tensor{Y}_{k}\Tensor{V}_{k}
	\label{Equation:Network:Admittance}
\end{equation}
where $\Tensor{Y}_{k}$ is the \emph{compound admittance matrix} \cite{Book:Arrillaga:1990:System}.



The state vector $\xa_{k}$ is composed of the voltage phasors $\Tensor{V}_{k}$
\begin{equation}
	\xa_{k}
	=	\left[
		\begin{array}{c}
			\Real{\Tensor{V}_{k}}\\
			\Imaginary{\Tensor{V}_{k}}
		\end{array}
		\right]
	\label{Equation:System:State:Vector}
\end{equation}
where $\Real{.}$ and $\Imaginary{.}$ denote the real and imaginary part.
So, there are $S=2|\Buses||\Phases|$ state variables in total.
The \DKF takes into account the statistical properties of the system whose state it estimates
using a linear \emph{process model} \cite{Journal:Kalman:1960:Estimation,Book:Milano:2016:Estimation}
\begin{equation}
	\xa_{k} = \Tensor{A}\xa_{k-1} + \Tensor{B}\ua_{k-1} + \nx_{k-1}
	\label{Equation:A}
\end{equation}
where $k\in\mathbb{N}$ is the index of the discrete time,
$\xa$ is the vector of state variables, $\ua$ is the vector of controllable variables, $\nx$ is the process noise, 
$\Tensor{A}$ links the system state at $k$ and $k-1$ in the absence of controllable variables and process noise, 
and $\Tensor{B}$ links the system state at $k$ with the controllable variables at $k-1$ in the absence of process noise.
For the case of a power system, the process model \eqref{Equation:A} can be simplified.
Firstly, \PMU[s] stream measurements at high refresh rates \cite{Standard:IEEE:2011:PMU:Transfer}
(typical refresh rates are in the order of tens of frames per second).
Therefore, there is only little variation in the state between any two consecutive time steps $k-1$ and $k$,
so that one may use a \emph{quasi-static} model with $\Tensor{A}=\Tensor{I}$.
Secondly, the inputs of a power system are not controllable from the point of view of the state estimator, 
and thus need not be considered in the process model.
Hence, one can set $\Tensor{B}=\Tensor{0}$.
Accordingly, \Equation{Equation:A} reduces to the well-known persistence process model
\begin{equation}
	\xa_{k} = \xa_{k-1} + \nx_{k-1}
	\label{Equation:System:State:Model}
\end{equation}
which is, equivalently, an \emph{Autoregressive Integrated Moving Average} (\ARIMA) model of order $(0,1,0)$.
This model has first been proposed for power transmission systems \cite{Journal:Debs:1970:Estimation:Linear:DKF},
but it also holds for power distribution systems as shown in \cite{Journal:Sarri:2016:SE:Linear:Performance},
where it is formally validated.
In particular, it is worthwhile noting that the process model can capture fast dynamics
if the associated time constants are reasonably longer than the time window used for the synchrophasor extraction,
i.e. several cycles of the fundamental component \cite{Journal:Belega:2013:PMU:Performance:IpDFT}%
\footnote
{%
	Typically, the window length is around $40$--$100$ milliseconds.
}.
Accordingly, slow transients with time constants of several hundred milliseconds can be treated,
while fast transients with time constants of a few tens of milliseconds cannot.
Namely, the fast transients are directly filtered by the \PMU measurements.



The measurement vector $\za_{k}$ is composed of nodal voltage phasors $\widetilde{\Tensor{V}}_{k}$
and nodal current phasors $\widetilde{\Tensor{I}}_{k}$, which are recorded at buses $\Instruments\subset\Buses$
that are equipped with \PMU[s].
Define the selector matrix $\Tensor{\Gamma}$ such that $\widetilde{\Tensor{V}}_{k}$ and $\widetilde{\Tensor{I}}_{k}$
may be expressed as
\begin{equation}
	\widetilde{\Tensor{V}}_{k} = \Tensor{\Gamma}\Tensor{V}_{k}
	~,~
	\widetilde{\Tensor{I}}_{k} = \Tensor{\Gamma}\Tensor{I}_{k}
	\label{Equation:B}
\end{equation}
In principle, different selector matrices could be chosen for mapping $\Tensor{V}_{k}$ to $\widetilde{\Tensor{V}}_{k}$
and $\Tensor{I}_{k}$ to $\widetilde{\Tensor{I}}_{k}$.
In practice, it is reasonable to assume that a \PMU measures voltage and current,
so the selector matrix is the same.
In analogy to the state vector $\xa_{k}$, the measurement vector $\za_{k}$ is defined in block form
\cite{Book:Milano:2016:Estimation}
\begin{equation}
	\za_{k}
	=	\left[
		\begin{array}{c}
			\Real{\widetilde{\Tensor{V}}_{k}}			\\
			\Imaginary{\widetilde{\Tensor{V}}_{k}}	\\
			\Real{\widetilde{\Tensor{I}}_{k}}			\\
			\Imaginary{\widetilde{\Tensor{I}}_{k}}
		\end{array}
		\right]
	\label{Equation:System:Measurement:Vector}
\end{equation}
Accordingly, there are in total $D=4|\Instruments||\Phases|$ measurements.
The \emph{measurement model}, which links the state vector $\xa_{k}$ with the measurement vector $\za_{k}$, 
is given by the linear equation
\begin{equation}
	\za_{k} = \Tz_{k}\xa_{k} + \nz_{k}
	\label{Equation:System:Measurement:Model}
\end{equation}
where $\nz_{k}$ is the measurement noise.
Use \eqref{Equation:Network:Admittance} and \eqref{Equation:B} to find
\begin{equation}
	\Tz_{k}
	=	\left[
		\begin{aligned}
				&\Tensor{\Gamma} 						&		&\Tensor{0}	\\
				&\Tensor{0}									&		&\Tensor{\Gamma}	\\
			+	&\Tensor{\Gamma}\Tensor{G}_{k}	&	-	&\Tensor{\Gamma}\Tensor{B}_{k}	\\
			+	&\Tensor{\Gamma}\Tensor{B}_{k}	&	+	&\Tensor{\Gamma}\Tensor{G}_{k}	\\
		\end{aligned}
		\right]
\end{equation}
where $\Tensor{G}_{k}=\Real{\Tensor{Y}_{k}}$ and $\Tensor{B}_{k}=\Imaginary{\Tensor{Y}_{k}}$.
In order for the system to be observable, the matrix $\Tz_{k}$ has to have full rank. 
\begin{Hypothesis}[Observability]
	\label{Hypothesis:Observability}
	The matrix $\Tz_{k}$ has full rank.
\end{Hypothesis}
In the following, it is always assumed that the placement of the \PMU[s] is done such that
this hypothesis holds \cite{Journal:Manousakis:2012:PMU:Placement:Taxonomy}.



The process noise $\nx_{k}$ and the measurement noise $\nz_{k}$ are modeled as 
spectrally white, zero-mean, normally distributed, and mutually uncorrelated random variables
\cite{Book:Simon:2006:SE}.
Formally
\begin{align}
	\nx_{k} &\sim \Normal{\Tensor{0},\Sx_{k}}
	\label{Equation:System:State:Noise}
	\\
	\nz_{k} &\sim \Normal{\Tensor{0},\Sz_{k}}
	\label{Equation:System:Measurement:Noise}
	\\
	\Sx_{k} &=	\Expectation{\nx_{k}\nx_{k}^{T}}
	\label{Equation:System:State:Covariance}
	\\
	\Sz_{k} &=	\Expectation{\nz_{k}\nz_{k}^{T}}
	\label{Equation:System:Measurement:Covariance}
	\\
	\Expectation{\nx_{k}\nz_{k}^{T}} &= \Tensor{0}
	\label{Equation:System:Uncorrelatedness}
\end{align}
where $\Normal{\Tensor{\mu},\Tensor{\Sigma}}$ designates the multivariate standard normal distribution
with mean vector $\Tensor{\mu}$ and covariance matrix $\Tensor{\Sigma}$,
and $\Expectation{.}$ denotes the expected value.
The process noise covariance matrix $\Sx_{k}$ is usually assumed to be diagonal, 
whereas the measurement noise covariance matrix $\Sz_{k}$ may be dense.
In the above measurement model, there is an implicit transformation from polar to rectangular coordinates,
since the \PMU[s] provide $\widetilde{\Tensor{V}}_{k}$ and $\widetilde{\Tensor{I}}_{k}$ in magnitude and phase,
whereas $\za_{k}$ is defined using real and imaginary parts \Equation{Equation:System:Measurement:Vector}.
It is important to note that this coordinate transformation does not substantially affect the normality
of the measurement error distribution in rectangular coordinates \Equation{Equation:System:Measurement:Noise}.
Indeed, it has recently been demonstrated in \cite{Dissertation:Sarri:2016:SE} that the normality is preserved
for practical values of the sensor accuracy in polar coordinates.
That is, the standard deviation of the measurement error would have to exceed $5\%$
for the effect to become noticeable (see \cite{Dissertation:Sarri:2016:SE} for further details).
Since \PMU[s] are typically equipped with voltage and current sensors with class $1$ or better,
\Equation{Equation:System:Measurement:Noise} holds in practice.
However, the coordinate transformation does affect the uncertainty associated with the measurements.
That is, the uncertainties associated with the rectangular coordinates
are a function of the uncertainties associated with the polar coordinates.
The interested reader is referred to \Appendix{Appendix:Transformation},
where this subject is illustrated in detail.

\subsection{The Discrete Kalman Filter}
\label{Section:Algorithm:Batch}



The \DKF estimates the state $\xa_{k}$ in two steps \cite{Book:Simon:2006:SE}.
First, an \emph{a priori} estimate $\xe_{k}^{-}$ is obtained using only the past measurements $\{\za_{l}:l<k\}$.
Thereafter, a refined \emph{a posteriori} estimate $\xe_{k}^{+}$ is computed
by considering all measurements $\{\za_{l}:l \leqslant k\}$ up to the present one.
Formally
\begin{align}
	\xe_{k}^{-}
	&=	\Expectation{\xa_{k}|\{\za_{l}:l<k\}}
	\label{Equation:Prediction:State}
	\\
	\xe_{k}^{+}
	&=	\Expectation{\xa_{k}|\{\za_{l}:l \leqslant k\}}
	\label{Equation:Estimation:State}
\end{align}
Henceforth, $\xa_{k}$ will be referred to as the \emph{true} state, $\xe_{k}^{-}$ as the \emph{predicted} state,
and $\xe_{k}^{+}$ as the \emph{estimated} state.
The prediction error $\e_{k}^{-}$ and the estimation error $\e_{k}^{+}$ are naturally defined as
\begin{align}
	\e_{k}^{-}	&=	\xa_{k}-\xe_{k}^{-}
	\label{Equation:Prediction:Error}
	\\
	\e_{k}^{+}	&=	\xa_{k}-\xe_{k}^{+}
	\label{Equation:Estimation:Error}
\end{align}
and the associated error covariance matrices are given by
\begin{align}
	\Se_{k}^{-}		&=	\Expectation{\e_{k}^{-}(\e_{k}^{-})^{T}}
	\label{Equation:Prediction:Covariance}
	\\
	\Se_{k}^{+}	&=	\Expectation{\e_{k}^{+}(\e_{k}^{+})^{T}}
	\label{Equation:Estimation:Covariance}
\end{align}



The objective for designing any \SE, including the Kalman Filter,
is to minimize the weighted norm of the estimation error
\begin{equation}
	\xe_{k}^{+} = \ArgMin\Expectation{(\e_{k}^{+})^{T}\Tensor{\Omega}_{k}\e_{k}^{+}}
	\label{Equation:C}
\end{equation}
where $\Tensor{\Omega}_{k}$ is a positive definite weighting matrix.
If $\nx_{k}$ and $\nz_{k}$ behave as described by 
\Equation{Equation:System:State:Noise}--\Equation{Equation:System:Uncorrelatedness},
then the \DKF is a solution of problem \Equation{Equation:C}, as shown in \cite{Book:Brown:2012:SE:Kalman}.
\begin{Algorithm}[Discrete Kalman Filter]
	\label{Algorithm:Batch}
	Consider a system described by a process model of the form \Equation{Equation:System:State:Model},
	and a measurement model of the form \Equation{Equation:System:Measurement:Model}
	that satisfies Hypothesis \ref{Hypothesis:Observability}.
	The \DKF can be formulated as follows (see \cite{Book:Simon:2006:SE}):
	\newline
	The prediction (\emph{a priori} estimation) step is defined by
	\begin{align}
		\xe_{k}^{-}
		&=	\xe_{k-1}^{+}
		\label{Equation:Batch:Prediction:State}
		\\
		\Se_{k}^{-}	
		&=	\Se_{k-1}^{+} + \Sx_{k}
		\label{Equation:Batch:Prediction:Covariance}
	\end{align}
	The estimation (\emph{a posteriori} estimation) step is defined by
	\begin{align}					
		\K_{k}
		&=	\Se_{k}^{-}\Tz_{k}^{T}(\Tz_{k}\Se_{k}^{-}\Tz_{k}^{T}+\Sz_{k})^{-1}
		\label{Equation:Batch:Estimation:Gain:A}				
		\\
		\xe_{k}^{+}	
		&=	\xe_{k}^{-} + \K_{k}(\za_{k}-\Tz_{k}\xe_{k}^{-})
		\label{Equation:Batch:Estimation:State:A}
		\\
		\Se_{k}^{+}
		&=	(\Tensor{I}-\K_{k}\Tz_{k})\Se_{k}^{-}
		\label{Equation:Batch:Estimation:Covariance:A}
	\end{align}
	which may alternatively be written as
	\begin{align}					
		(\Se_{k}^{+})^{-1}
		&=	(\Se_{k}^{-})^{-1} + \Tz_{k}^{T}\Sz_{k}^{-1}\Tz_{k}
		\label{Equation:Batch:Estimation:Covariance:B}
		\\
		\K_{k}
		&=	\Se_{k}^{+}\Tz_{k}^{T}\Sz_{k}^{-1}
		\label{Equation:Batch:Estimation:Gain:B}	
		\\
		\xe_{k}^{+}	
		&=	\xe_{k}^{-} + \K_{k}(\za_{k}-\Tz_{k}\xe_{k}^{-})
		\label{Equation:Batch:Estimation:State:B}
	\end{align}
	where $\K_{k}$ is the so-called \emph{Kalman Gain}.
\end{Algorithm}
Concerning the above, there are a few important comments to be made.
Firstly, in order for the \DKF to work properly, one must ensure that the term $\Tz_{k}\Se_{k}^{-}\Tz_{k}^{T}+\Sz_{k}$ 
in \Equation{Equation:Batch:Estimation:Gain:A} is invertible, respectively that $\Se_{k}^{-}$ and $\Se_{k}^{+}$
in \Equation{Equation:Batch:Estimation:Covariance:B} are invertible.
Most works in the literature tacitly assume that $\Se_{k}^{-}$ and $\Se_{k}^{+}$ are positive definite, 
which ensures that the aforementioned invertibility conditions hold.
Namely, if $\Se_{k}^{-}$ and $\Se_{k}^{+}$ are positive definite, they are invertible, too.
Since $\Tz_{k}$ has full rank by assumption, and $\Sz_{k}$ is positive semidefinite by definition,
it follows directly that $\Tz_{k}\Se_{k}^{-}\Tz_{k}^{T}+\Sz_{k}$ is also positive definite.
However, strictly speaking, covariance matrices are only guaranteed to be \emph{positive semidefinite},
not \emph{strictly positive definite}.	
For the sake of rigor, the assumption of strict positive definiteness is explicitly stated as a working hypothesis in this work.
\begin{Hypothesis}[Positive Definite Estimation Error Covariance]
	\label{Hypothesis:Definiteness}
	The estimation error covariance matrices $\Se_{k}^{-}$ and $\Se_{k}^{+}$
	are strictly positive definite (and therefore invertible).
\end{Hypothesis}
A motivation for why this hypothesis is indeed reasonable in practice is given in \Appendix{Appendix:Invertibility}.
In the following, it is always assumed that this working hypothesis holds.
Secondly, it is important to note that the above-stated alternative formulations
of the estimation step are indeed equivalent.
Since this property will be used later on during the proof of equivalence of the \DKF and the \SDKF,
it is explicitly stated in the following.
\begin{Lemma}[Equivalent \DKF Formulations]
	Provided that Hypothesis \ref{Hypothesis:Definiteness} holds, 
	the formulations \Equation{Equation:Batch:Estimation:Gain:A}--\Equation{Equation:Batch:Estimation:Covariance:A}
	and \Equation{Equation:Batch:Estimation:Covariance:B}--\Equation{Equation:Batch:Estimation:State:B}
	of the estimation step are equivalent.
\end{Lemma}
A proof can for instance be found in \cite{Book:Simon:2006:SE}.
Lastly, one should be aware of the fact that $\Sx_{k}$ influences the estimation accuracy of the \DKF.
Usually, it is assumed to be constant ($\Sx_{k}=\Sx$), and set to a value which ensures reasonable performance
for typically encountered dynamics.
Nevertheless, there are ways to assess $\Sx_{k}$ online in order to improve the accuracy.
For instance, it can be approximated as the sample variance of the estimates $\xe_{k}^{+}$
over a sliding time window \cite{Conference:Zanni:2014:SE:DKF:Covariance},
or computed formally by solving a $\log(\det(.))$ optimization problem \cite{Journal:Zanni:2016:SE:DKF:Covariance}.
However, such techniques are beyond the scope of this paper,
and are therefore not considered in the following.
Rather, the traditional approach of using a constant value is followed.

\subsection{The Sequential Discrete Kalman Filter}
\label{Section:Algorithm:Sequential}



In view of an implementation into dedicated hardware, the most critical operation is the matrix inversion,
because it cannot be parallelized and hence scales poorly.
Since all the involved operands depend on the time $k$,
the inversion has to be computed online in real-time, 
which emphasizes the need for a more efficient algorithm.
The estimation process can be simplified considerably, if it may be assumed that
the measurement noise variables $(\nz_{k})_{i}$ are mutually uncorrelated.
\begin{Hypothesis}[Uncorrelated Measurement Noise]
	\label{Hypothesis:Uncorrelatedness}
	The measurement noise variables $(\nz_{k})_{i}$ are mutually uncorrelated, 
	so the measurement noise covariance matrix $\Sz_{k}$ is diagonal
	\begin{equation}
		(\Sz_{k})_{ij}
		=
		\left\{
		\begin{array}{cc}
			\sigma_{i}^{2}		&	(i = j)		\\
			0						&	(i \neq j)
		\end{array}
		\right.
		\label{Equation:System:Noise:Diagonality}
	\end{equation}
	where $\sigma_{i}$ denotes the standard deviation of $(\nz_{k})_{i}$.
\end{Hypothesis}
One should note that this is not a strong assumption.
Indeed, the impact of measurement correlation on state estimator performance in power distribution systems
has for instance been investigated in \cite{Journal:Muscas:2014:SE:Correlation}.
In this study, the correlation factors inferred for commercial \PMU installations have been found to be so low,
that the estimation accuracy cannot be improved when they are considered in the measurement model.
Even for a hypothetical experiment with very high correlation factors,
no noteworthy improvement in estimation accuracy has been observed.
Finally, it is worth observing that \cite{Journal:Muscas:2014:SE:Correlation} considers measurements in polar coordinates, 
whereas this work uses rectangular coordinates as stated in \Equation{Equation:System:Measurement:Vector}.
Since the transformation from polar to rectangular coordinates does not affect the normality
of the measurement error distribution, as it has been explained in \Section{Section:Algorithm:Model}, 
the findings of \cite{Journal:Muscas:2014:SE:Correlation} do still apply.
Therefore, it is justified to assume that Hypothesis \ref{Hypothesis:Uncorrelatedness} holds.
In this case, the \SDKF can be used instead of the \DKF.


\begin{Algorithm}[Sequential Discrete Kalman Filter]
	\label{Algorithm:Sequential}
	Consider a system described by a process model of the form \Equation{Equation:System:State:Model},
	and a measurement model of the form \Equation{Equation:System:Measurement:Model}
	that satisfies Hypotheses \ref{Hypothesis:Observability} and \ref{Hypothesis:Uncorrelatedness}.
	The \SDKF can be formulated as follows (see \cite{Book:Simon:2006:SE}):
	\newline	
	The prediction (\emph{a priori} estimation) step is defined by
	\begin{align}
		\xe_{k}^{-}
		&=	\xe_{k-1}^{+}
		\label{Equation:Sequential:Prediction:State}
		\\
		\Se_{k}^{-}
		&=	\Se_{k-1}^{+} + \Sx_{k-1}
		\label{Equation:Sequential:Prediction:Covariance}
	\end{align}
	The estimation (\emph{a posteriori} estimation) step treats the elements of $\za_{k}$ sequentially.
	Using the index $i\in\{1,\ldots,D\}$ for $\za_{k}$, the individual measurement $\za_{k,i}$,
	its measurement model $\Tz_{k,i}$, and its measurement noise covariance $\Sz_{k,i}$ are defined as
	\begin{align}
		\za_{k,i}		&=	(\za_{k})_{i}
		\label{Equation:Sequential:Notation:Measurement}
		\\
		\Tz_{k,i}		&=	\row_{i}(\Tz_{k})
		\label{Equation:Sequential:Notation:Link}
		\\
		\Sz_{k,i}		&=	(\Sz_{k})_{ii}
		\label{Equation:Sequential:Notation:Noise}
	\end{align}
	Set the initial values $\xe_{k,0}^{+}$ and $\Se_{k,0}^{+}$ to
	\begin{align}
		\xe_{k,0}^{+}
		&=	\xe_{k}^{-}
		\label{Equation:Sequential:Initial:State}
		\\
		\Se_{k,0}^{+}
		&=	\Se_{k}^{-}
		\label{Equation:Sequential:Initial:Covariance}
	\end{align}
	Compute $\xe_{k,i}^{+}$, and $\Se_{k,i}^{+}$ sequentially for $i\in\{1,\ldots,D\}$
	\begin{align}
		\K_{k,i}
		&=	\Se_{k,i-1}^{+}\Tz_{k,i}^{T}(\Tz_{k,i}\Se_{k,i-1}^{+}\Tz_{k,i}^{T}+\Sz_{k,i})^{-1}
		\label{Equation:Sequential:Estimation:Gain:A}
		\\
		\xe_{k,i}^{+}
		&=	\xe_{k,i-1}^{+} + \K_{k,i}(\za_{k,i}-\Tz_{k,i}\xe_{k,i-1}^{+})
		\label{Equation:Sequential:Estimation:State:A}
		\\
		\Se_{k,i}^{+}
		&=	(\Tensor{I}-\K_{k,i}\Tz_{k,i})\Se_{k,i-1}^{+}
		\label{Equation:Sequential:Estimation:Covariance:A}
	\end{align}
	or alternatively using
	\begin{align}
		(\Se_{k,i}^{+})^{-1}
		&=	(\Se_{k,i-1}^{+})^{-1} + \Tz_{k,i}^{T}\Sz_{k,i}^{-1}\Tz_{k,i}
		\label{Equation:Sequential:Estimation:Covariance:B}
		\\
		\K_{k,i}
		&=	\Se_{k,i}^{+}\Tz_{k,i}^{T}\Sz_{k,i}^{-1}
		\label{Equation:Sequential:Estimation:Gain:B}
		\\
		\xe_{k,i}^{+}
		&=	\xe_{k,i-1}^{+} + \K_{k,i}(\za_{k,i}-\Tz_{k,i}\xe_{k,i-1}^{+})
		\label{Equation:Sequential:Estimation:State:B}
	\end{align}	
	The final results $\xe_{k}^{+}$ and $\Se_{k}^{+}$ are obtained after $D$ iterations
	\begin{align}
		\xe_{k}^{+}		&=	\xe_{k,D}^{+}
		\label{Equation:Sequential:Final:State}
		\\
		\Se_{k}^{+}	&=	\Se_{k,D}^{+}
		\label{Equation:Sequential:Final:Covariance}
	\end{align}
\end{Algorithm}
Observe that the equations describing the estimation step of the \SDKF
are similar to those of the \DKF.
Analogously
\begin{Lemma}[Equivalent \SDKF Formulations]
	Provided that Hypotheses \ref{Hypothesis:Definiteness} and \ref{Hypothesis:Uncorrelatedness} hold, the two formulations
	\Equation{Equation:Sequential:Estimation:Gain:A}--\Equation{Equation:Sequential:Estimation:Covariance:A}
	and \Equation{Equation:Sequential:Estimation:Covariance:B}--\Equation{Equation:Sequential:Estimation:State:B}
	of the estimation step are equivalent.
\end{Lemma}
The proof for the \DKF in \cite{Book:Simon:2006:SE} applies with minor changes.

\subsection{Proof of Equivalence}
\label{Section:Algorithm:Equivalence}


%
\begin{Theorem}[Equivalence of \DKF and \SDKF]
	Consider a system defined by a process model of the form \Equation{Equation:System:State:Model},
	and a measurement model of the form \Equation{Equation:System:Measurement:Model}
	that fulfils Hypothesis \ref{Hypothesis:Observability}.
	If Hypotheses \ref{Hypothesis:Definiteness} and \ref{Hypothesis:Uncorrelatedness} hold,
	the \DKF as given in Algorithm \ref{Algorithm:Batch} and the \SDKF as given in Algorithm \ref{Algorithm:Sequential}
	are equivalent.
\end{Theorem}
Although the \SDKF does appear in the literature (e.g. \cite{Book:Simon:2006:SE,Book:Brown:2012:SE:Kalman}),
to the best of the authors' knowledge, a formal proof of equivalence is nowhere to be found.
Therefore, it is now proven that the \DKF and the \SDKF are indeed equivalent.
Since the prediction equations are clearly identical, it suffices to show that the estimation equations yield the same results.



\begin{Proof}[Equivalence of $\Se_{k}^{+}$]
	Note that \Equation{Equation:Sequential:Estimation:Covariance:B} defines $(\Se_{k,i}^{+})^{-1}$ recursively.
	Expand the recursion for $(\Se_{k,D}^{+})^{-1}$ to obtain
	\begin{equation}
		(\Se_{k,D}^{+})^{-1} = (\Se_{k,0}^{+})^{-1} + \sum_{i=1}^{D}\Tz_{k,i}^{T}\Sz_{k,i}^{-1}\Tz_{k,i}
	\end{equation}
	Since $\Sz_{k}$ is diagonal according to \Equation{Equation:System:Noise:Diagonality},
	where $\Sz_{k,i}=(\Sz_{k})_{ii}$ are the diagonal elements, and $\Tz_{k,i}=\row_{i}(\Tz_{k})$, it follows
	\begin{align}
		\sum_{i=1}^{D}\Tz_{k,i}^{T}\Sz_{k,i}^{-1}\Tz_{k,i}
		&=	\sum_{i=1}^{D}\row_{i}^{T}(\Tz_{k})(\Sz_{k})_{ii}^{-1}\row_{i}(\Tz_{k})
		\\
		&=	\Tz_{k}^{T}\Sz_{k}\Tz_{k}
	\end{align}
	Use $\Se_{k}^{-}=\Se_{k,0}^{+}$ from \Equation{Equation:Sequential:Initial:Covariance},
	and $\Se_{k}^{+}=\Se_{k,D}^{+}$ from \Equation{Equation:Sequential:Final:Covariance} to find
	\begin{equation}
		(\Se_{k}^{+})^{-1} = (\Se_{k}^{-})^{-1} + \Tz_{k}^{T}\Sz_{k}\Tz_{k}
	\end{equation}
	Obviously, this is identical to \Equation{Equation:Batch:Estimation:Covariance:B} of the \DKF,
	which proves the part of the claim concerning $\Se_{k}^{+}$.
	\QEDtotal
\end{Proof}



The proof of equivalence of $\xe_{k}^{+}$ involves some chain terms
that are produced by the unraveling of the sequential computation.
To keep the equations concise, the ordered matrix chain product $\Product$
with decreasing index is defined here for later use
\begin{equation}
	\Product^{i=m}_{n}(\Tensor{M}_{i})
	=	\Tensor{M}_{m} \times \Tensor{M}_{m-1} \times \ldots \times \Tensor{M}_{n+1} \times \Tensor{M}_{n}
\end{equation}

\begin{Proof}[Equivalence of $\xe_{k}^{+}$]
	Group the terms in \Equation{Equation:Sequential:Estimation:State:B} with respect to
	the estimated state $\xe_{k,i}^{+}$ and the measurement $\za_{k,i}$
	\begin{equation}
		\xe_{k,i}^{+} = (\Tensor{I}-\K_{k,i}\Tz_{k,i})\xe_{k,i-1}^{+} + \K_{k,i}\za_{k,i}
	\end{equation}
	Obviously, this defines $\xe_{k,i}^{+}$ recursively.
	Expand the recursion for $\xe_{k,D}^{+}$, and group the terms with respect to $\xe_{k,0}^{+}$ and $\za_{k}$
	\begin{equation}
		\xe_{k,D}^{+} = \Tensor{\psi}_{k} + \Tensor{\varphi}_{k}
		\label{Equation:Equivalence:State}
	\end{equation}
	where the group terms $\Tensor{\psi}_{k}$ and $\Tensor{\varphi}_{k}$ are given by
	\begin{align}
		\Tensor{\psi}_{k}
		&=	\Product^{j=D}_{1}\left\{\Tensor{I}-\K_{k,j}\Tz_{k,j}\right\}\xe_{k,0}^{+}
		\\
		\Tensor{\varphi}_{k}
		&=	\K_{k,D}\za_{k,D} + \sum_{i=1}^{D-1}\Product^{j=D}_{i+1}\left\{\Tensor{I}-\K_{k,j}\Tz_{k,j}\right\}\K_{k,i}\za_{k,i}
	\end{align}
	For \Equation{Equation:Equivalence:State} and \Equation{Equation:Batch:Estimation:State:B} to be equivalent,
	it must hold that
	\begin{align}
		\Tensor{\psi}_{k}		&=	(\Tensor{I}-\K_{k}\Tz_{k})\xe_{k,0}^{+}
		\label{Equation:Equivalence:Prediction}
		\\
		\Tensor{\varphi}_{k}	&=	\K_{k}\za_{k}
		\label{Equation:Equivalence:Measurement}
	\end{align}
	which will be proven in the following.\\
	%
	%
	\emph{Proof ($\Tensor{\psi}_{k}$)}.
	Remember that the equivalence has already been proven for $\Se_{k}^{+}$.
	Therefore, the recursive formula \Equation{Equation:Sequential:Estimation:Covariance:A}
	gives the same results as \Equation{Equation:Batch:Estimation:Covariance:A}
	after $D$ iterations.
	It follows that
	\begin{equation}
		\Product_{1}^{j=D}\left\{\Tensor{I}-\K_{k,j}\Tz_{k,j}\right\}\Se_{k,0}^{+} = (\Tensor{I}-\K_{k}\Tz_{k})\Se_{k}^{-}
	\end{equation}
	Recall that $\Se_{k,0}^{+}=\Se_{k}^{-}$ from \Equation{Equation:Sequential:Initial:Covariance}, 
	so obviously
	\begin{equation}
		\Product_{1}^{j=D}\left\{\Tensor{I}-\K_{k,j}\Tz_{k,j}\right\} = \Tensor{I}-\K_{k}\Tz_{k}
	\end{equation}
	Multiplying each side of the above equation by $\xe_{k,0}^{+}$ produces $\Tensor{\psi}_{k}$
	on the left-hand side, which proves claim \Equation{Equation:Equivalence:Prediction}.
	\QEDpartial\newline
	%
	%
	\emph{Proof ($\Tensor{\varphi}_{k}	$)}.
	Solve \Equation{Equation:Sequential:Estimation:Covariance:A}
	for the term $(\Tensor{I}-\K_{k,i}\Tz_{k,i})$ to obtain
	\begin{equation}
		(\Tensor{I}-\K_{k,i}\Tz_{k,i}) = \Se_{k,i}^{+}(\Se_{k,i-1}^{+})^{-1}
	\end{equation}
	From the above, it follows straightforward that
	\begin{align}
		\Product^{j=D}_{i+1}\left\{\Tensor{I}-\K_{k,j}\Tz_{k,j}\right\}
		&=	\Product^{j=D}_{i+1}\left\{\Se_{k,j}^{+}(\Se_{k,j-1}^{+})^{-1}\right\}\\
		&=	\Se_{k,D}^{+}(\Se_{k,i}^{+})^{-1}
	\end{align}
	Substitute this into the definition of $\Tensor{\varphi}_{k}$, which yields
	\begin{equation}
		\Tensor{\varphi}_{k} = \K_{k,D}\za_{k,D} + \Se_{k,D}^{+}\sum_{i=1}^{D-1}(\Se_{k,i}^{+})^{-1}\K_{k,i}\za_{k,i}
	\end{equation}
	Since $\K_{k,i}=\Se_{k,i}^{+}\Tz_{k,i}^{T}\Sz_{k,i}^{-1}$ according to \Equation{Equation:Sequential:Estimation:Gain:B},
	it follows that
	\begin{equation}
		\Tensor{\varphi}_{k} = \Se_{k,D}^{+}\sum_{i=1}^{D}\Tz_{k,i}^{T}\Sz_{k,i}^{-1}\za_{k,i}
	\end{equation}
	As $\Sz_{k}$ is diagonal with elements $\Sz_{k,i}=(\Sz_{k})_{ii}$ \Equation{Equation:System:Noise:Diagonality},
	and $\Tz_{k,i}=\row_{i}(\Tz_{k})$ \Equation{Equation:Sequential:Notation:Link}, this may be rewritten as
	\begin{align}
		\sum_{i=1}^{D}\Tz_{k,i}^{T}\Sz_{k,i}^{-1}\za_{k,i}
		&=	\sum_{i=1}^{D}\row_{i}^{T}(\Tz_{k})(\Sz_{k})_{ii}^{-1}(\za_{k})_{i}\\
		&=	\Tz_{k}^{T}\Sz_{k}^{-1}\za_{k}
	\end{align}
	Use the above and the fact that $\Se_{k,D}^{+}=\Se_{k}^{+}$, as already proven,
	to simplify the expression for $\Tensor{\varphi}_{k}$, namely
	\begin{equation}
		\Tensor{\varphi}_{k} = \Se_{k}^{+}\Tz_{k}^{T}\Sz_{k}^{-1}\za_{k}
	\end{equation}
	Since the gain is defined as $\K_{k}=\Se_{k}^{+}\Tz_{k}^{T}\Sz_{k}^{-1}$ in \Equation{Equation:Batch:Estimation:Gain:B},
	it becomes apparent that the claim \Equation{Equation:Equivalence:Measurement} indeed holds.
	\QEDpartial\newline
	%
	%
	Having verified that the claims \Equation{Equation:Equivalence:Prediction}
	and \Equation{Equation:Equivalence:Measurement} hold,
	it follows that the obtained $\xe_{k}^{+}$ is indeed identical for both filters.
	\QEDtotal
\end{Proof}

\subsection{Computational Complexity}
\label{Section:Algorithm:Complexity}

{
\renewcommand{\arraystretch}{1.5}
\begin{table}[t]
	\caption{Computational Complexity (\DKF)}
	\label{Table:Batch:Overall}		
	\centering
	\begin{tabular}{lcc}
		\hline
		\textbf{Prediction}		&$+|-$								&$\times|\div$	\\
		\hline
		$\xe_{k}^{-}$				&$0$									&$0$					\\
		$\Se_{k}^{-}$				&$S$									&$0$					\\
		\hline
		\textbf{Estimation}		&$+|-$								&$\times|\div$	\\
		\hline
		$\dTz_{k}$					&$DS(S-1)$						&$DS^{2}$			\\
		$\K_{k}$					&$2D^{2}S+D(1-D-S)+m$	&$2D^{2}S+n$	\\
		$\xe_{k}^{+}$				&$2DS$							&$2DS$			\\
		$\Se_{k}^{+}$				&$DS^{2}$							&$DS^{2}$			\\
		\hline
	\end{tabular}
\end{table}
\begin{table}[t]
	\caption{Computational Complexity (\SDKF)}
	\label{Table:Sequential:Overall}		
	\centering
	\begin{tabular}{lccc}
		\hline
		\multicolumn{2}{l}{\textbf{Prediction}}	&$+|-$				&$\times|\div$	\\
		\hline
		$\xe_{k}^{-}$		&								&$0$					&$0$					\\
		$\Se_{k}^{-}$		&								&$S$					&$0$					\\
		\hline
		\multicolumn{2}{l}{\textbf{Estimation}}	&$+|-$				&$\times|\div$	\\
		\hline
		$\dTz_{k,i}$		&$i\in\{1,\ldots,D\}$		&$DS(S-1)$		&$DS^{2}$			\\
		$\K_{k,i}$			&ditto						&$DS$				&$D(2S+1)$		\\
		$\xe_{k,i}^{+}$	&ditto						&$2DS$			&$2DS$			\\
		$\Se_{k,i}^{+}$	&ditto						&$DS^{2}$			&$DS^{2}$			\\
		\hline
	\end{tabular}
\end{table}
}



It is important to note that the formulation
\Equation{Equation:Sequential:Estimation:Gain:A}--\Equation{Equation:Sequential:Estimation:Covariance:A}
does not feature a matrix inversion.
Recall that $\Tz_{k,i}=\row_{i}(\Tz_{k})$ is a row vector \Equation{Equation:Sequential:Notation:Link},
and that $\Sz_{k,i}=(\Sz_{k})_{ii}$ is a scalar \Equation{Equation:Sequential:Notation:Noise}.
Therefore, the term $\Tz_{k,i}\Se_{k,i-1}^{+}\Tz_{k,i}^{T}+\Sz_{k,i}$ is also a scalar.
Moreover, the \SDKF using formulation
\Equation{Equation:Sequential:Estimation:Gain:A}--\Equation{Equation:Sequential:Estimation:Covariance:A}
requires fewer operations than the \DKF using formulation 
\Equation{Equation:Batch:Estimation:Gain:A}--\Equation{Equation:Batch:Estimation:Covariance:A}.
Tables \ref{Table:Batch:Overall} and \ref{Table:Sequential:Overall} summarize the computational complexity
of the \DKF and the \SDKF, respectively.
In \Appendix{Appendix:Complexity}, this aspect is further analyzed with respect to elementary operations
in Tables \ref{Table:Batch:Detailed} and \ref{Table:Sequential:Detailed} for deeper insight.
Note that the terms $m\in\Order{D^{3}}$ and $n\in\Order{D^{3}}$ scale with $D^{3}$.



Investigating Tables \ref{Table:Batch:Overall} and \ref{Table:Sequential:Overall} reveals that the \SDKF and the \DKF
only differ in the amount of operations invested into the computation of $\K_{k}$ and $\K_{k,i}$ ($i\in\{1,\ldots,D\}$),
respectively.
Clearly, the \SDKF needs fewer operations than the \DKF, since
\begin{align}
	(+|-)				&	&DS 			&<	2D^{2}S+D(1-D-S)+m	\\
	(\times|\div)	&	&D(2S+1) 	&<	2D^{2}S+n
\end{align}
In particular, the \SDKF lacks the cubic terms $m,n\in\Order{D^{3}}$, which stem from the matrix inversion
(see \Table{Table:Sequential:Detailed}).
In order for the system to be observable, it is a necessary condition that the number of measurements
be equal to or larger than the number of states, that is $D\geq S$
(recall that a sufficient condition is given in Hypothesis \ref{Hypothesis:Observability}).
For the sake of security, one usually ensures that there is ample measurement redundancy,
which means that $D\gg S$.
In such a case, the matrix inversion limits the performance of the \DKF, 
because a very large matrix needs to be inverted.
Conversely, the \SDKF only requires basic matrix-vector operations and some scalar divisions
(see \Table{Table:Sequential:Detailed}).
In contrast to the matrix inversion, these operations are rather simple, 
and may hence be implemented in dedicated hardware.
Moreover, they can be parallelized to accelerate the computation.
In conclusion, the \SDKF is suitable for an \FPGA implementation,
whereas the \DKF is not.


\section{Hardware Implementation}
\label{Section:Implementation}



A \NI CompactRIO microcontroller is used for the implementation,
more precisely a \NI-\CRIO-9033 \cite{Specification:NI:2014:CRIO:9033}.
This device is equipped both with an \FPGA (Xilinx Kintex-7 7K160T) and a \CPU (Intel Atom E3825),
and can therefore host both the \emph{Model Under Test} (\MUT) and the \emph{Testbench} (\TB).
The prototype implementation of the \SDKF-\SE is discussed in \Section{Section:Implementation:Prototype},
and \TB setup is described in \Section{Section:Implementation:Testbench}.


\subsection{Prototype}
\label{Section:Implementation:Prototype}


\begin{figure}[t]
	\centering
	\includegraphics[width=1.0\linewidth]{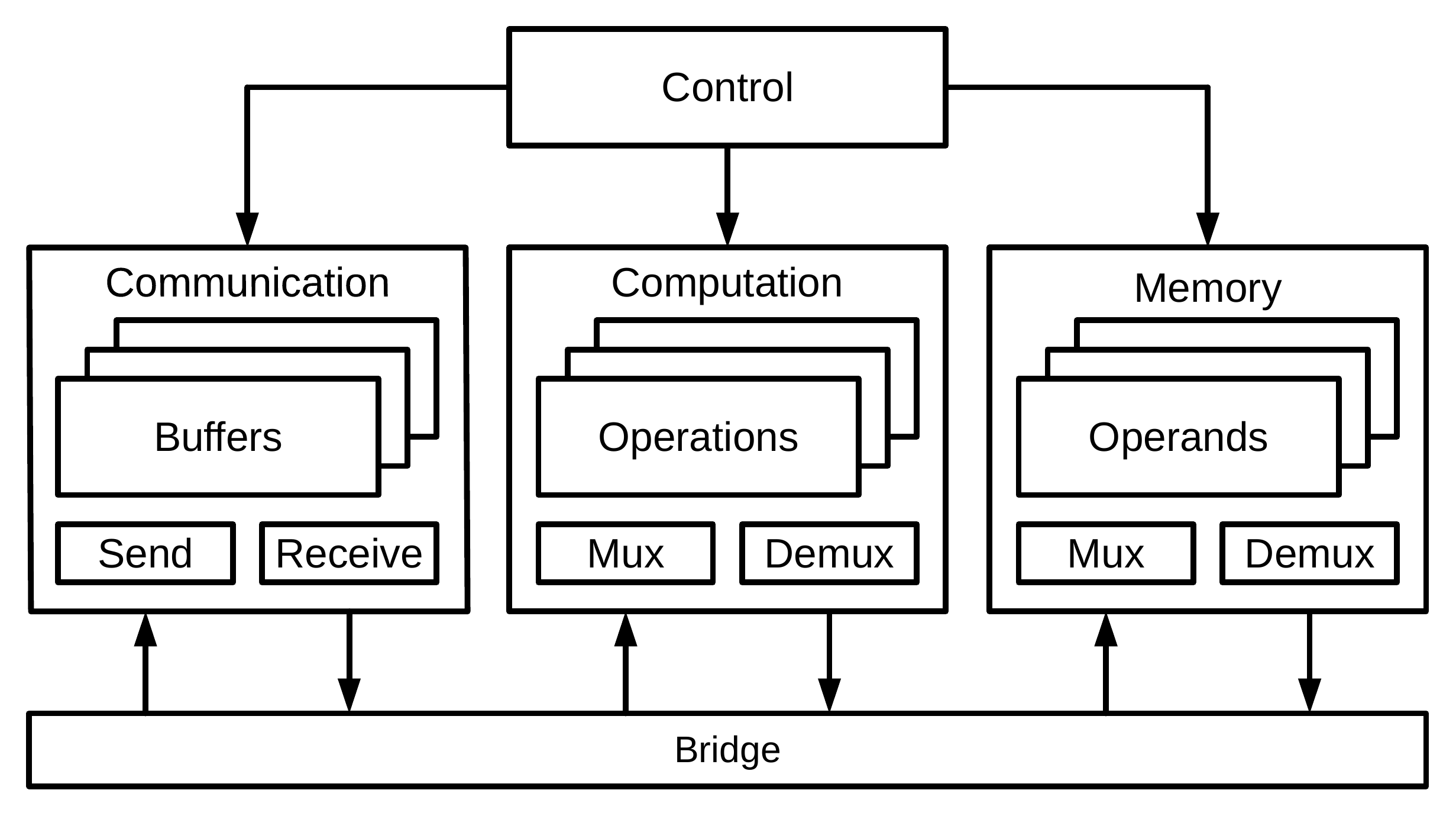}
	\caption{Division of the architecture into specialized modules.}
	\label{Figure:Modules}
\end{figure}

\Figure{Figure:Modules} shows the division of the architecture into modules for
\emph{communication}, \emph{computation}, \emph{memory}, and \emph{control}.



The \emph{communication} module manages the exchange of data between the \CPU and the \FPGA.
For this purpose, \emph{First-In First-Out} (\FIFO) buffers implemented in the on-chip
\emph{Random Access Memory} (\RAM) of the \FPGA are used.
The transfer process itself is managed by a \emph{Direct Memory Access} (\DMA) controller on the low-level,
and coordinated by a handshake protocol using interrupts on the high-level.



\begin{figure}[t]
	\centering
	\includegraphics[width=0.7\linewidth]{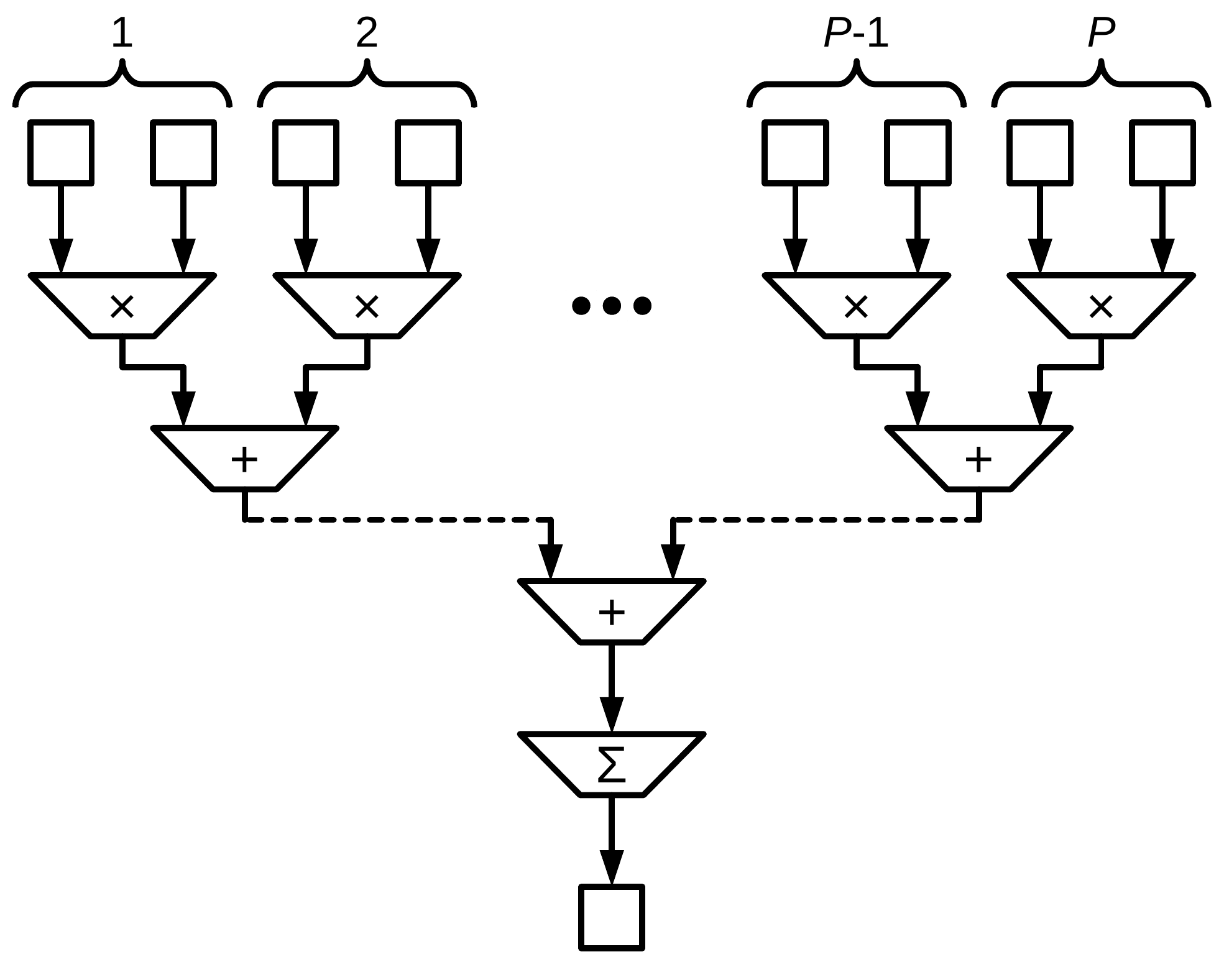}
	\caption{Parallelized implementation of the inner product.}
	\label{Figure:Dot}
\end{figure}

The \emph{computation} module comprises all resources for the actual calculations.
According to \Table{Table:Sequential:Detailed}, the following operations are needed:
(i) matrix addition / subtraction ($\Tensor{M}_{1}\pm\Tensor{M}_{2}$),
(ii) vector addition / subtraction ($\Tensor{v}_{1}\pm\Tensor{v}_{2}$) and scaling ($s\cdot\Tensor{v}$),
(iii) the outer product ($\Tensor{v}_{1}\Tensor{v}_{2}^{T}$), 
(iv) the inner product ($\Tensor{v}_{1}^{T}\Tensor{v}_{2}$), and
(v) the matrix-vector product ($\Tensor{M}\Tensor{v}$).
In order to achieve high throughput, these operations are pipelined and parallized.	
Since parallel processing requires parallel data access,
the operands need to be partitioned into blocks and stored in separate memories.
Recall that the \SDKF processes the measurements sequentially,
so the parallelization is done with respect to the states.
Say $P$ the degree of parallelization,
then the matrix operands ($\Se_{k}^{-}$, $\Se_{k,i}^{+}$) are split into rasters of $P \times P$ blocks,
and the vector operands ($\Tz_{k,i}$, $\dTz_{k,i}$, $\K_{k,i}$, $\xe_{k}^{-}$, $\xe_{k,i}^{+}$) into arrays of $P$ blocks.
Accordingly, the operations (i), (iii), and (v) are sped up by a factor of $P^{2}$,
whereas (ii) and (iv) are accelerated by a factor of $P$.
Of course, this requires the allocation of a corresponding number of arithmetic blocks.
Note that the operations (i)--(iii) are straightforward to parallelize using arrays of adders or multipliers.
The inner product (iv) can be built from a multiplier array, an adder tree, and one accumulator
as depicted in \Figure{Figure:Dot}.
The matrix-vector product (v) is in turn made from $P$ replicas of (iv).
For the synthesis of the arithmetic blocks, optimized libraries for \emph{Single-Precision Floating-Point} (\SGL) operations
which exploit the \emph{Digital Signal Processing} (\DSP) slices of the \FPGA
to achieve high performance \cite{Specification:XI:2016:DSP}, are used.
When configuring each block, a trade-off has to be made between throughput, latency, and resource consumption.
For the \RTSE application, high throughput and low resource consumption are crucial.
The resulting configuration is listed in \Table{Table:Operations}.

{
\renewcommand{\arraystretch}{1.5}
\begin{table}[t]
	\caption{Configuration of the arithmetic blocks}
	\label{Table:Operations}
	\centering
	\begin{tabular}{ccrc}
		\hline
		Operation		&Throughput 		&\multicolumn{1}{c}{Latency}	&\DSP[s]	\\	
		\hline
		$\pm$			&$1$ / cycle		&$5$ cycles							&$2$			\\	
		$\times$		&$1$ / cycle		&$2$ cycles							&$3$			\\	
		$\sum$			&$1$ / cycle		&$20$ cycles							&$9$			\\	
		$\div$			&$1$ / cycle		&$20$ cylces							&$8$			\\	
		\hline
	\end{tabular}	
\end{table}
}



The \emph{memory} module contains the storage for the operands.
%
%
Note that one does not need to store all the intermediate results listed in \Table{Table:Sequential:Detailed}.
Indeed, some of these operations are contracted in the \FPGA implementation to increase the performance.
Hence, it suffices to store $\Sx_{k}$, $\Sz_{k}$, $\Tz_{k}$, $\za_{k}$, $\dTz_{k,i}$, $\K_{k,i}$,
$\xe_{k}$ ($\xe_{k,i}^{+}$ / $\xe_{k}^{-}$), $\Se_{k}$ ($\Se_{k,i}^{+}$ / $\Se_{k}^{-}$), $\W_{k,i}^{-1}$ and $\ze_{k,i}$.
Recall from the above discussion that these operands are partitioned into blocks,
which need to be stored in separate memories to allow for parallel processing.
Therefore, one needs to take into consideration both the size and the organization of the available \RAM
when selecting the degree of parallelization $P$ for a given hardware platform.
Firstly, there has to be enough memory (in terms of bits) to house the operands as a whole.
Secondly, there need to be enough separate \RAM slices for distributing the operands,
which are divided into $P \times P$ or $P$ blocks, respectively. 


{
\renewcommand{\arraystretch}{1.5}
\begin{table}[t]
	\caption{Resource occupation}
	\label{Table:Resources}
	\centering
	\begin{tabular}{lrrc}
		\hline
		Resource		&Available 	&Occupied		&Percentage	\\
		\hline
		\FF[s]			&202'800		&49'088			&24.2	\\
		\LUT[s]			&101'400		&43'166			&42.6	\\
		\DSP[s]			&600				&357				&59.5	\\
		\RAM[s]			&325				&262				&80.6	\\
		\hline
	\end{tabular}
\end{table}
}



With the \FPGA resources available on the \NI-\CRIO-9033,
the degree of parallelization that can be achieved with this architecture is $P=4$.
\Table{Table:Resources} lists the resource occupation in terms of \emph{Flip-Flops} (\FF[s]), \emph{Look-Up Tables} (\LUT[s]),
\DSP[s], and \RAM[s] obtained for this value of $P$.
Clearly, the \DSP[s] and the \RAM[s] are the most critical resources.
The high number of \DSP[s] required is mainly due to the operations that require $P \times P$ arrays of arithmetic blocks,
namely the matrix addition or subtraction (i), the outer product (iii), and the matrix-vector product (v).
The high utilization of \RAM[s] is mostly due to the operands $\Se_{k}$ and $\Tz_{k}$,
which are matrices whose number of elements is proportional to the square of the network size.
The \FF[s] and \LUT[s] are principally used as shift registers for pipelining, 
but are obviously not critical resources.

\subsection{Testbench}
\label{Section:Implementation:Testbench}


\begin{figure}[t]
	\centering
	\includegraphics[width=0.9\linewidth]{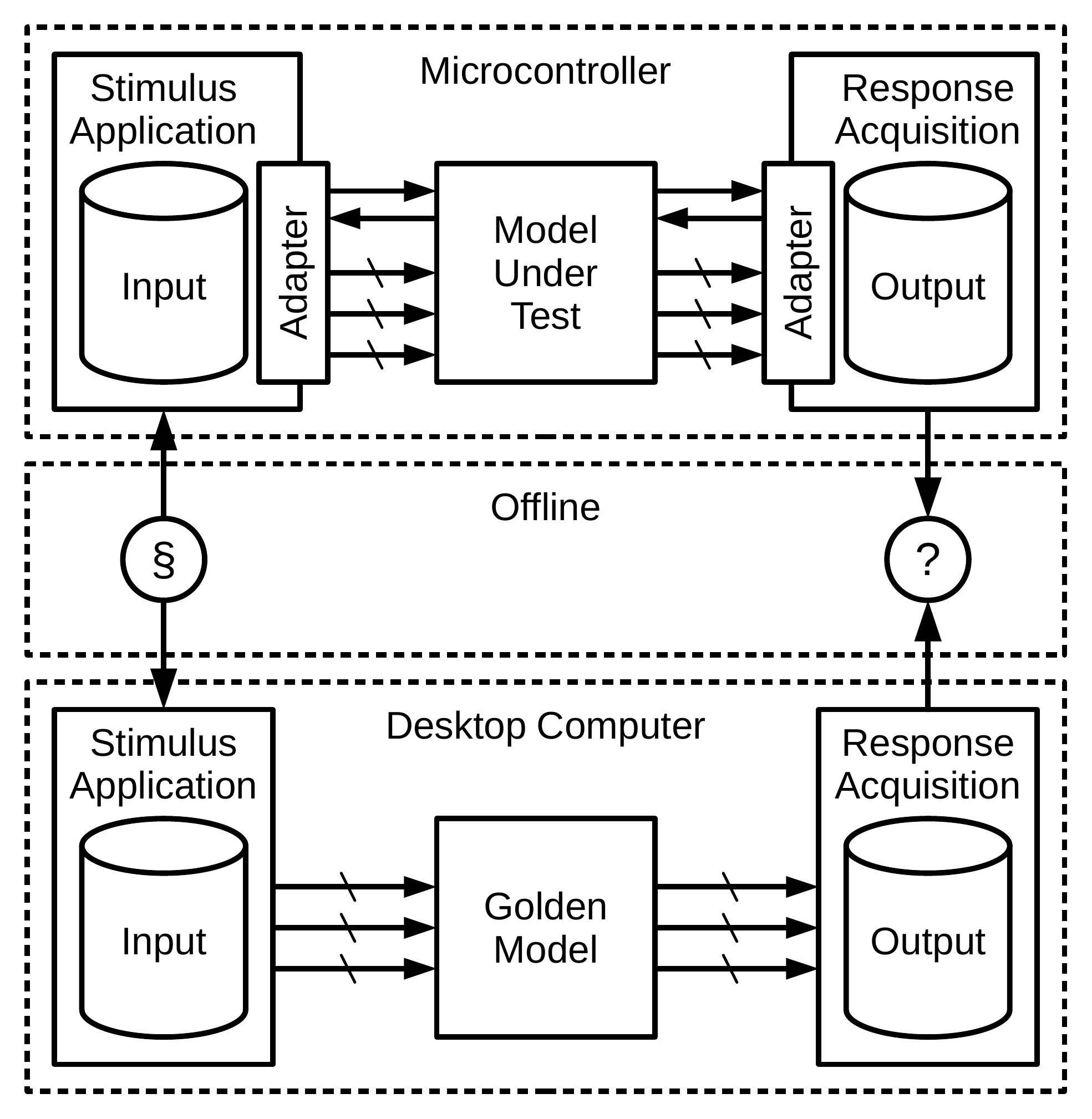}
	\caption{Schematic of the \TB setup.}
	\label{Figure:Testbench}
\end{figure}

The \TB setup depicted in \Figure{Figure:Testbench} is used to validate the hardware implementation.
It is divided into two separate parts associated to the \MUT and the \emph{Golden Model} (\GM),
which serves as the reference for the validation.
The \MUT part comprises the \FPGA implementation of the \SDKF along with some \CPU software,
and is executed on the \NI-\CRIO-9033, which runs \texttt{NI Linux Real-Time} and \texttt{NI LabVIEW}.
The \CPU software fulfills two purposes.
Firstly, it coordinates the communication with the \FPGA by handshaking,
and steers the \DMA controller that manages the data transfer.
Secondly, it provides IO functionality for the \TB files, namely reading the stimuli and writing the responses.
In particular, there are \emph{protocol adapters} which abstract the interface between
the high-level data of the \TB and the low-level data of the \MUT.
The \GM part consists of a \texttt{MATLAB} implementation of the \DKF,
and is executed on a desktop machine under \texttt{Mac OSX}.
Since the stimuli and the responses are stored in files,
the \MUT and the \GM may be run independently.
Therefore, the validation of the responses can be done offline.

\section{Experimental Validation}
\label{Section:Validation}


This section is dedicated to the validation of the developed hardware prototype.
First, the results of the functional verification, which is based on test data for a benchmark distribution feeder,
are presented in \Section{Section:Validation:Functionality}.
Then, the results of a scalability analysis, which is conducted using random data,
are discussed in \Section{Section:Validation:Scalability}.


\subsection{Functional Verification}
\label{Section:Validation:Functionality}



{
\renewcommand{\arraystretch}{1.5}

\begin{table}[t!]
	\centering
	\caption{Removed nodes}
	\label{Table:Modification}
	\begin{tabular}{cc}
		\hline
		Type		&	Nodes (Naming according to \cite{Journal:Kersting:1991:Feeder:Distribution})	\\
		\hline
		Tie	 	&	802, 806, 808, 812, 818, 824, 854, 858, 834, 836 			\\
		\hline
	\end{tabular}
\end{table}

\begin{table}[t!]
	\centering
	\caption{Distributed generation and load}
	\label{Table:Injection}
	\begin{tabular}{cc}
		\hline
		Type		&	Nodes (Naming according to \cite{Journal:Kersting:1991:Feeder:Distribution})	\\
		\hline
		\DG		&	822, 856, 848, 838														\\
		\DL		&	810, 816, 820, 826, 828, 832, 890, 864, 844, 860, 840	\\
		\hline
	\end{tabular}
\end{table}

\begin{table}[t!]
	\centering
	\caption{PMU placement}
	\label{Table:PMU}
	\begin{tabular}{cc}
		\hline
		Type		&	Nodes (Naming according to \cite{Journal:Kersting:1991:Feeder:Distribution})	\\
		\hline
		PMU		&	800, 806, 810, 816, 820, 822, 826, 828, 836	\\
					&	832, 890, 864, 844, 848, 860, 840, 830			\\
		\hline
	\end{tabular}
\end{table}
}

The benchmark system used for the functional verification is adapted from the \IEEE 34-node distribution test feeder
\cite{Journal:Kersting:1991:Feeder:Distribution},
which is an unbalanced three-phase grid with a rated line-to-line voltage of $24.9\,\kV$ (RMS).
The per unit base is chosen as $V_{b}=24.9\,\kV$ and $S_{b}=1\,\MVA$.
For this work, the original configuration given in \cite{Journal:Kersting:1991:Feeder:Distribution} is modified slightly
by removing very short lines connected in series with very long ones (through merge).
Note that the resulting reduced network is electrically equivalent to the original one,
but does not consider some of its nodes with null injections (see \Table{Table:Modification}).
The distribution feeder is connected to the feeding subtransmission grid in node 800.
This link is characterized by a short-circuit power of $S_{sc}=300\,\MVA$,
and a short-circuit impedance $Z_{sc}$ with $R_{sc}/X_{sc}=0.1$.
Furthermore, it is assumed that the voltage behind $Z_{sc}$ of the subtransmission system is constant,
which implies that the corresponding feeding node behaves as an ideal slack.
The lines are unbalanced and made from the same type of cable, so the per-unit-length resistance $\Tensor{R}'$,
reactance $\Tensor{X}'$, and susceptance $\Tensor{B}'$ are identical for all lines.
These parameters are listed in detail in \cite{Book:Milano:2016:Estimation}.
Both generation and load are distributed across the entire feeder, as listed in \Table{Table:Injection}.
The profiles stem from a measurement campaign conducted on the \EPFL campus
in Lausanne, Switzerland \cite{Conference:Pignati:2015:SE:DKF:Demonstrator}.
Hence, the \emph{distributed load} (\DL) is a composition of offices and workshops,
and the \emph{distributed generation} (\DG) are photovoltaic panels, which only inject active power.
\begin{figure}[t]
	\centering
	
	\subfloat[Aggregated nodal injections (\DG).]
	{
		\centering
		\label{Figure:Injection:Generation}
		\includegraphics[width=0.98\linewidth]{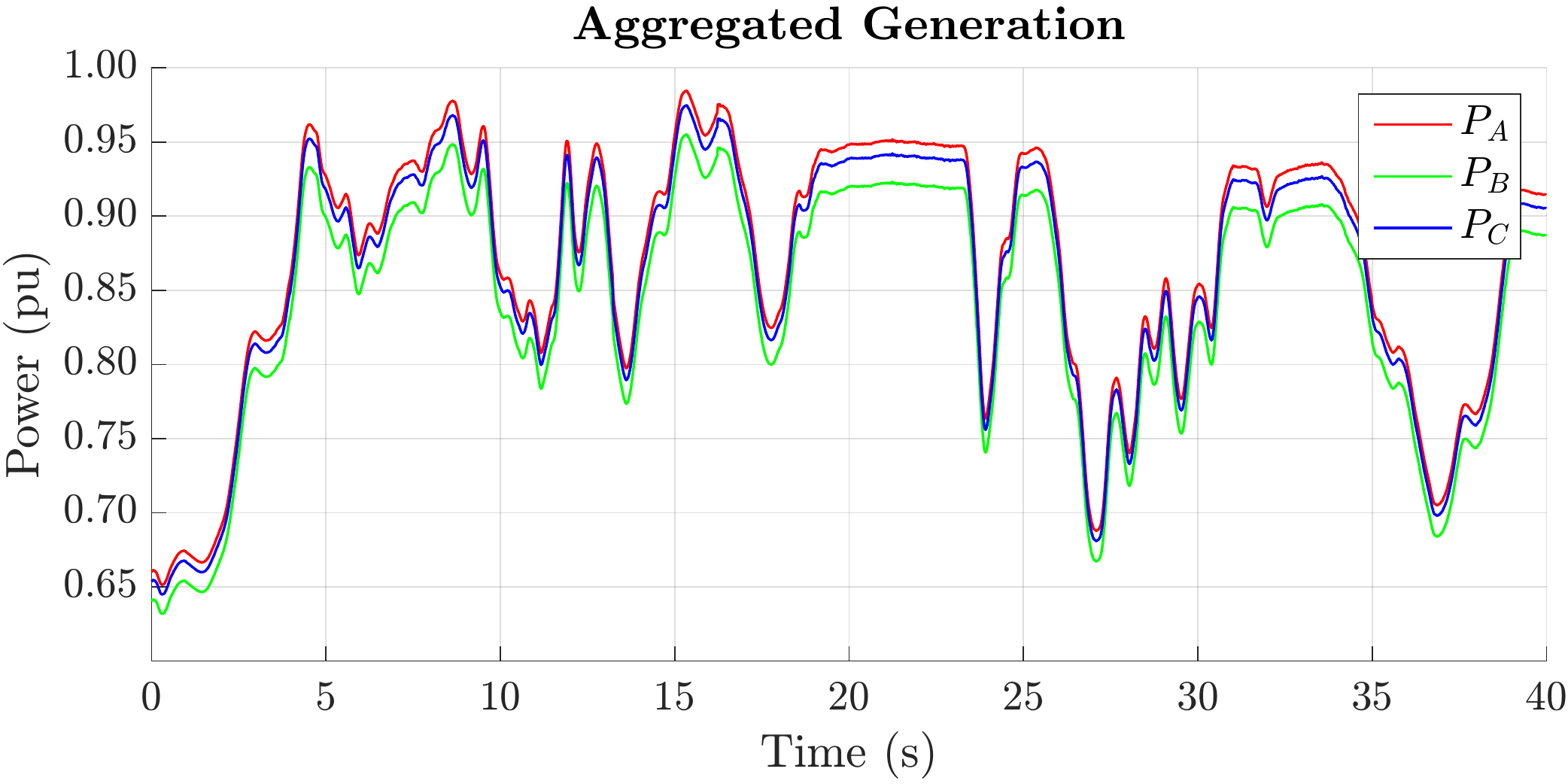}
	}
	
	\subfloat[Aggregated nodal absorptions (\DL).]
	{
		\centering
		\includegraphics[width=0.98\linewidth]{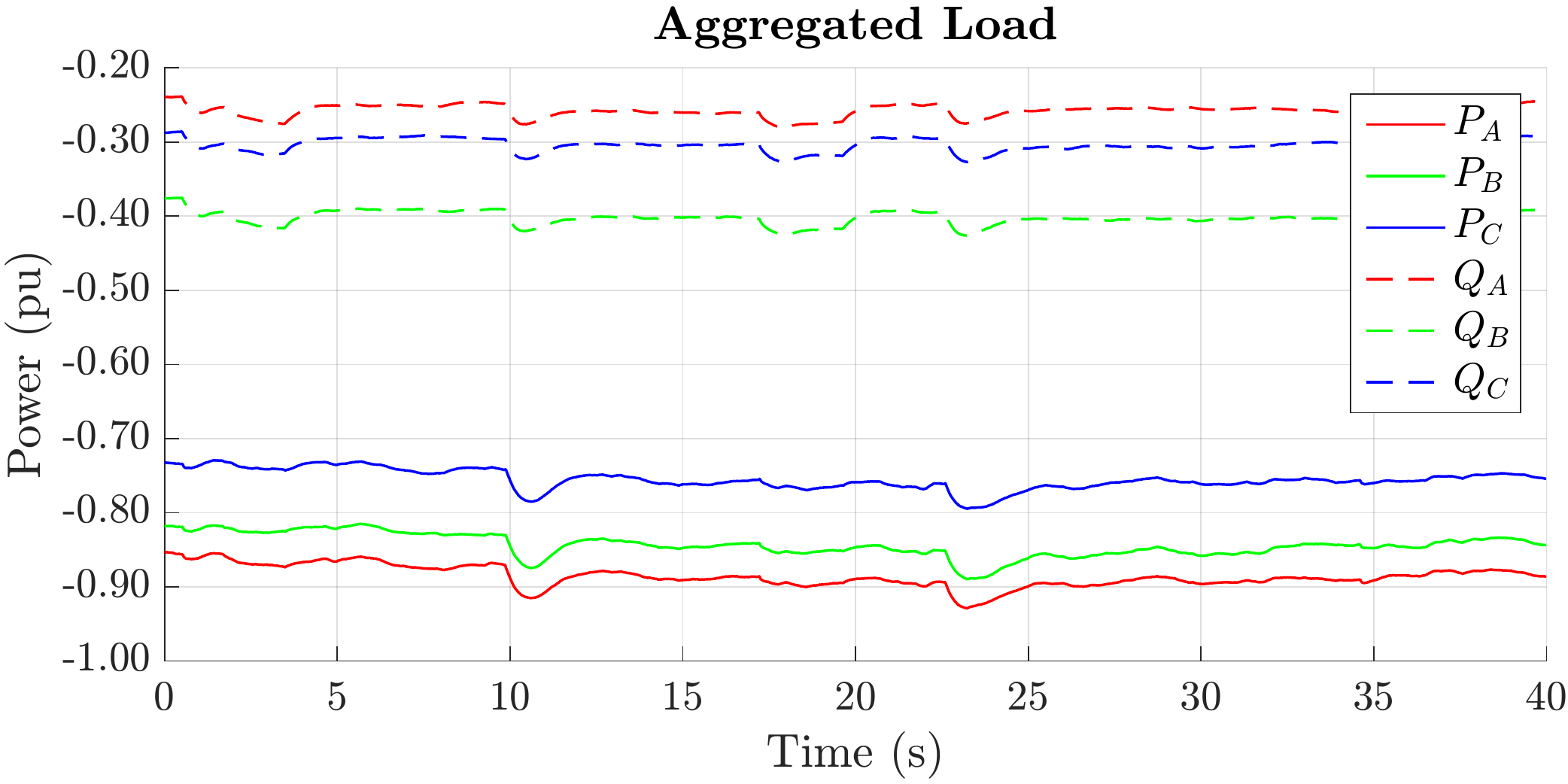}	
	}
	
	\caption{Aggregated power profiles.}
	\label{Figure:Injection}
\end{figure}
See \Figure{Figure:Injection} for the aggregated profiles of power injection and absorption
(generation is positive, load is negative).
The \PMU[s] are placed as given in \Table{Table:PMU} so that the system is observable,
which is ensured if $\Tz_{k}$ has full rank \cite{Journal:Manousakis:2012:PMU:Placement:Taxonomy}.
Each \PMU records the synchrophasors of nodal voltage and current in all phases
at a refresh rate of 50 frames per second.
The measurement system of the \PMU[s] consists of class $0.1$ / $0.2$ voltage and current sensors
(see \cite{Standard:IEC:2010:Instrument:General,Standard:IEC:2014:Instrument:Current}).



\begin{figure}
	\centering
	\begin{algorithmic}[1]
		\Procedure{Measurements}{$\Tensor{S}_{k}$, $\Tensor{Y}_{k}$, $e_{\rho}$, $e_{\varphi}$}
			\State $\Tensor{V}_{k} \leftarrow \Call{LoadFlow}{\Tensor{S}_{k},\Tensor{Y}_{k}}$
			\State $\widetilde{\Tensor{V}}_{k}=\Call{AddPolarNoise}{\Tensor{V}_{k},e_{\rho},e_{\varphi}}$
			\State $\Tensor{I}_{k}=\Tensor{Y}_{k}\Tensor{V}_{k}$
			\State $\widetilde{\Tensor{I}}_{k}=\Call{AddPolarNoise}{\Tensor{I}_{k},e_{\rho},e_{\varphi}}$
			\State \Return $\Tensor{V}_{k}$, $\widetilde{\Tensor{V}}_{k}$, $\widetilde{\Tensor{I}}_{k}$
		\EndProcedure
		\State
		\Procedure{AddPolarNoise}{$\Tensor{x}$, $e_{\rho}$, $e_{\varphi}$}
			\State $\sigma_{\rho}\leftarrow(1/3)\cdot e_{\rho}$
			\State $\sigma_{\varphi}\leftarrow(1/3)\cdot e_{\varphi}$
			\For {$i\in\{1,\ldots,\Call{Length}{\Tensor{x}}\}$}
				\State $\rho\leftarrow(1+\Call{Gauss}{0,\sigma_{\rho}})\cdot|\Tensor{x}(i)|$
				\State $\varphi\leftarrow(1+\Call{Gauss}{0,\sigma_{\phi}})\cdot\angle(\Tensor{x}(i))$				
				\State $\widetilde{\Tensor{x}}(i)\leftarrow\rho\angle\varphi$
			\EndFor
			\State \Return $\widetilde{\Tensor{x}}$
		\EndProcedure
	\end{algorithmic}
	\caption{Pseudocode describing the test data preparation.}
	\label{Figure:Preparation}
\end{figure}

To prepare the test data, the procedure in \Figure{Figure:Preparation} is followed.
The admittance matrix $\Tensor{Y}_{k}$ and the nodal powers $\Tensor{S}_{k}$
define a  \emph{Load Flow} (\LF) problem at each time $k$,
whose solution are the true nodal voltages $\Tensor{V}_{k}$.
In this respect, it is assumed that the topology of the network and the electrical parameters of the cables
do not change during the considered period of time, so $\Tensor{Y_{k}}=\Tensor{Y}$ is constant%
\footnote
{%
	If topological changes take place, the estimation process needs to be redone with the updated $\Tensor{Y}$
	(of the new topology) and a new initial state vector.
	One possible initialization is the so-called \emph{flat start} with nodal voltages equal to $1\PerUnit$
	and phase angle differences with respect to the slack equal to zero.
}.
The measurement accuracy is determined by the metrological characteristics of the \PMU[s] and their sensors.
In practice, the impact of the sensors on the accuracy dominates.
Thus, the measurements $\widetilde{\Tensor{V}}_{k}$ and $\widetilde{\Tensor{I}}_{k}$
may be obtained by perturbing the true values $\Tensor{V}_{k}$ and $\Tensor{I}_{k}$ with noise,
whose distribution is determined by the sensor properties (see \Figure{Figure:Preparation}).
For the used class $0.1$ / $0.2$ sensors, the inferred maximum errors are $e_{\rho}=10^{-3}~\PerUnit$ for magnitude
and $e_{\varphi}=1.5\cdot10^{-3}~\Radian$ for phase (see \cite{Journal:Sarri:2016:SE:Linear:Performance}).
One may reasonably suppose that the sensor performance does not change with time,
so $\Sz_{k}=\Sz$ is constant.
Recall from \Section{Section:Algorithm:Model} that $\Sz$ models the measurement uncertainties in rectangular coordinates.
The derivation of $\Sz$ from the uncertainties in polar coordinates is explained
in \Appendix{Appendix:Transformation}.
	
As previously explained in \Section{Section:Algorithm:Batch}, there exist online assessment methods for $\Sx_{k}$,
but they are beyond the scope of this paper.
For the sake of brevity, the process noise covariance matrix is assumed to be a constant diagonal matrix $\Sx_{k}=\Sx$
with all diagonal entries set to $10^{-6}~\PerUnit^{2}$.
Finally, the estimator needs initial values $\Se_{0}^{+}$ and $\xe_{0}^{+}$.
One can use $\Se_{0}^{+}=\Sx$, and set $\xe_{0}^{+}$ to a flat voltage profile.
Then, the responses $\xe_{k}^{+}|_\text{\GM}$ and $\xe_{k}^{+}|_\text{\MUT}$,
i.e. the estimated state $\xe_{k}^{+}$ provided by the \GM and the \MUT, are recorded in the \TB setup.
The corresponding estimated nodal voltage phasors $\Estimate{\Tensor{V}}_{k}|_\text{\MUT}$
and $\Estimate{\Tensor{V}}_{k}|_\text{\GM}$ are defined by \Equation{Equation:System:State:Vector}.



For the validation, one needs to look at the \emph{estimation accuracy} and the \emph{numerical accuracy}.
The former is related to the estimation error $\Estimate{\Tensor{V}}_{k}-\Tensor{V}_{k}$, where $\Estimate{\Tensor{V}}_{k}$
is a placeholder for $\Estimate{\Tensor{V}}_{k}|_\text{\MUT}$ and $\Estimate{\Tensor{V}}_{k}|_\text{\GM}$.
The latter corresponds to the mismatch $\Estimate{\Tensor{V}}_{k}|_\text{\MUT}-\Estimate{\Tensor{V}}_{k}|_\text{\GM}$.
To be more precise, one is interested in the statistical distribution of these quantities.
For this analysis, the true voltages $V_{b,p,k}$ and the estimated voltages $\Estimate{V}_{b,p,k}$
are expressed in polar coordinates.
That is
\begin{align}
	V_{b,p,k}					&=	|V_{b,p,k}|\angle\delta_{b,p,k}	\\
	\Estimate{V}_{b,p,k}	&=	|\Estimate{V}_{b,p,k}|\angle\Estimate{\delta}_{b,p,k}
\end{align}
Recall that $b\in\Buses$ is the bus, and $p\in\Phases$ is the phase.



\begin{figure}
	
	\subfloat[Magnitude error $|\Estimate{V}_{b,p,k}|-|V_{b,p,k}|$.]
	{%
		\centering
		\includegraphics[width=1.0\linewidth]{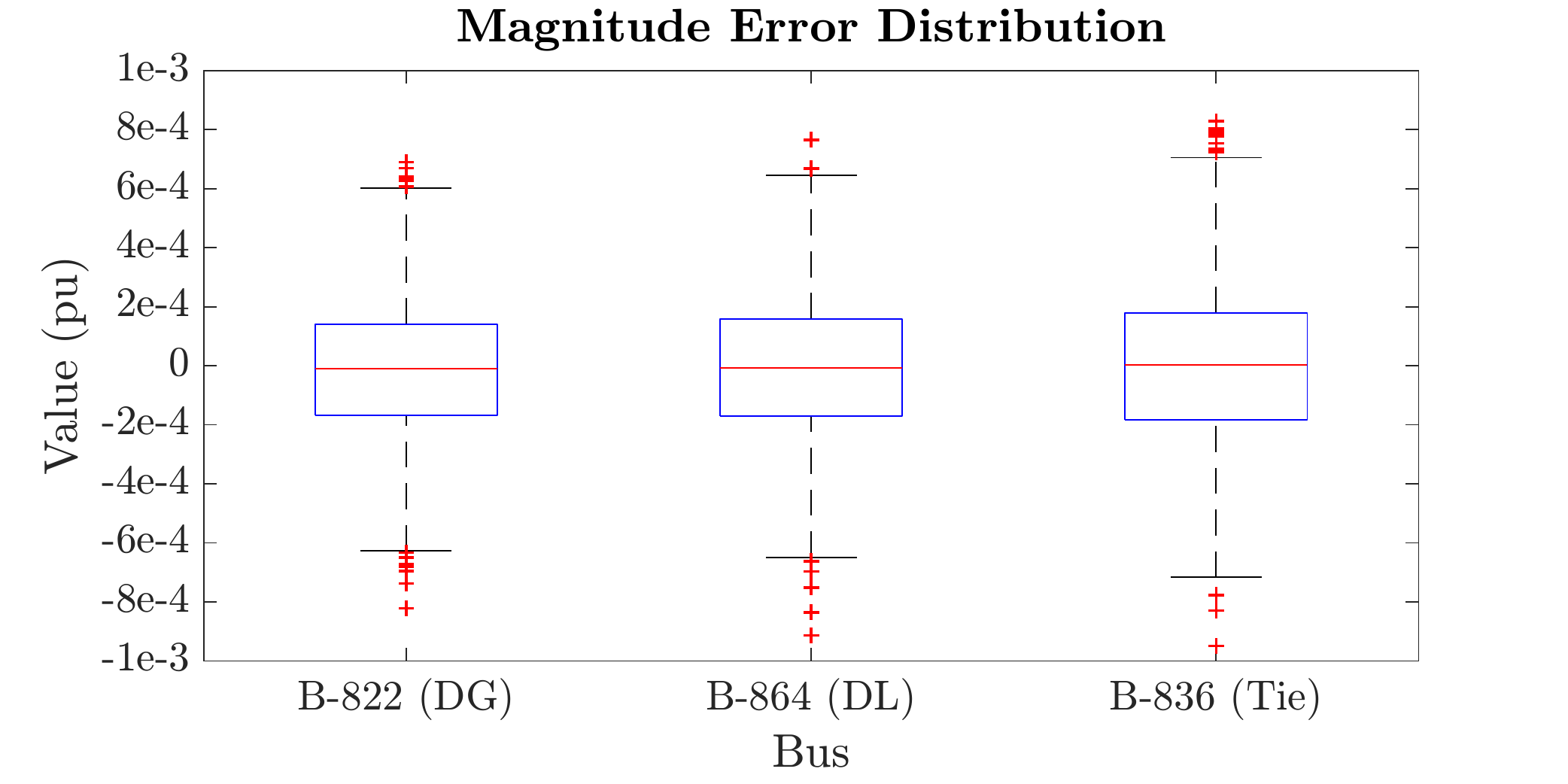}
		\label{Figure:Error:Magnitude}
	}
	
	\subfloat[Phase error $\Estimate{\delta}_{b,p,k}-\delta_{b,p,k}$.]
	{%
		\centering
		\includegraphics[width=1.0\linewidth]{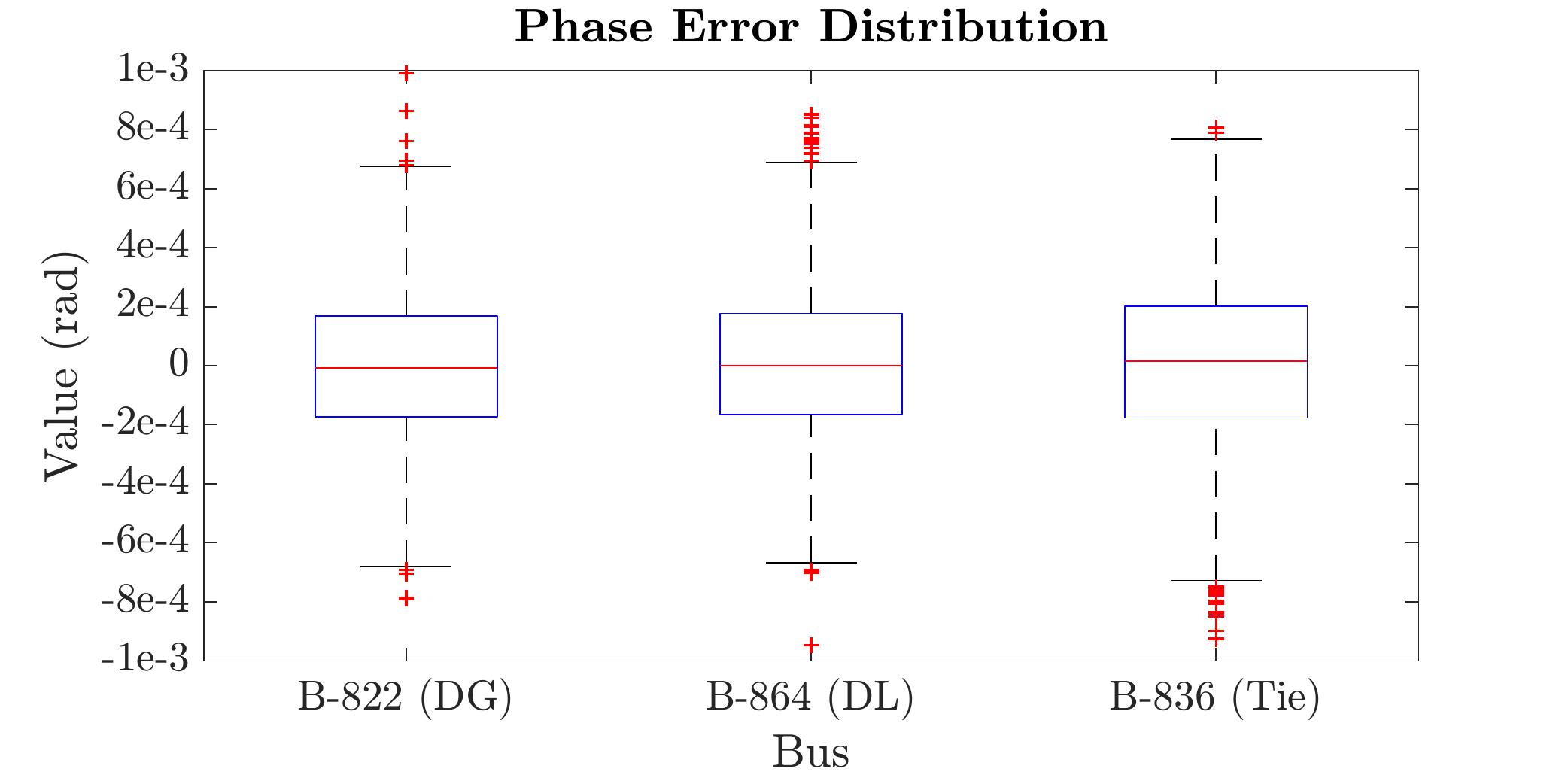}
		\label{Figure:Error:Phase}
	}
	
	\caption{Distribution of the error $\Estimate{V}_{b,p,k}-V_{b,p,k}$.}
	\label{Figure:Error}
\end{figure}

\begin{figure}
	
	\subfloat[Magnitude mismatch $|\Estimate{V}_{b,p,k}|_{\text{MUT}}-|\Estimate{V}_{b,p,k}|_{\text{GM}}$.]
	{%
		\centering	
		\includegraphics[width=1.0\linewidth]{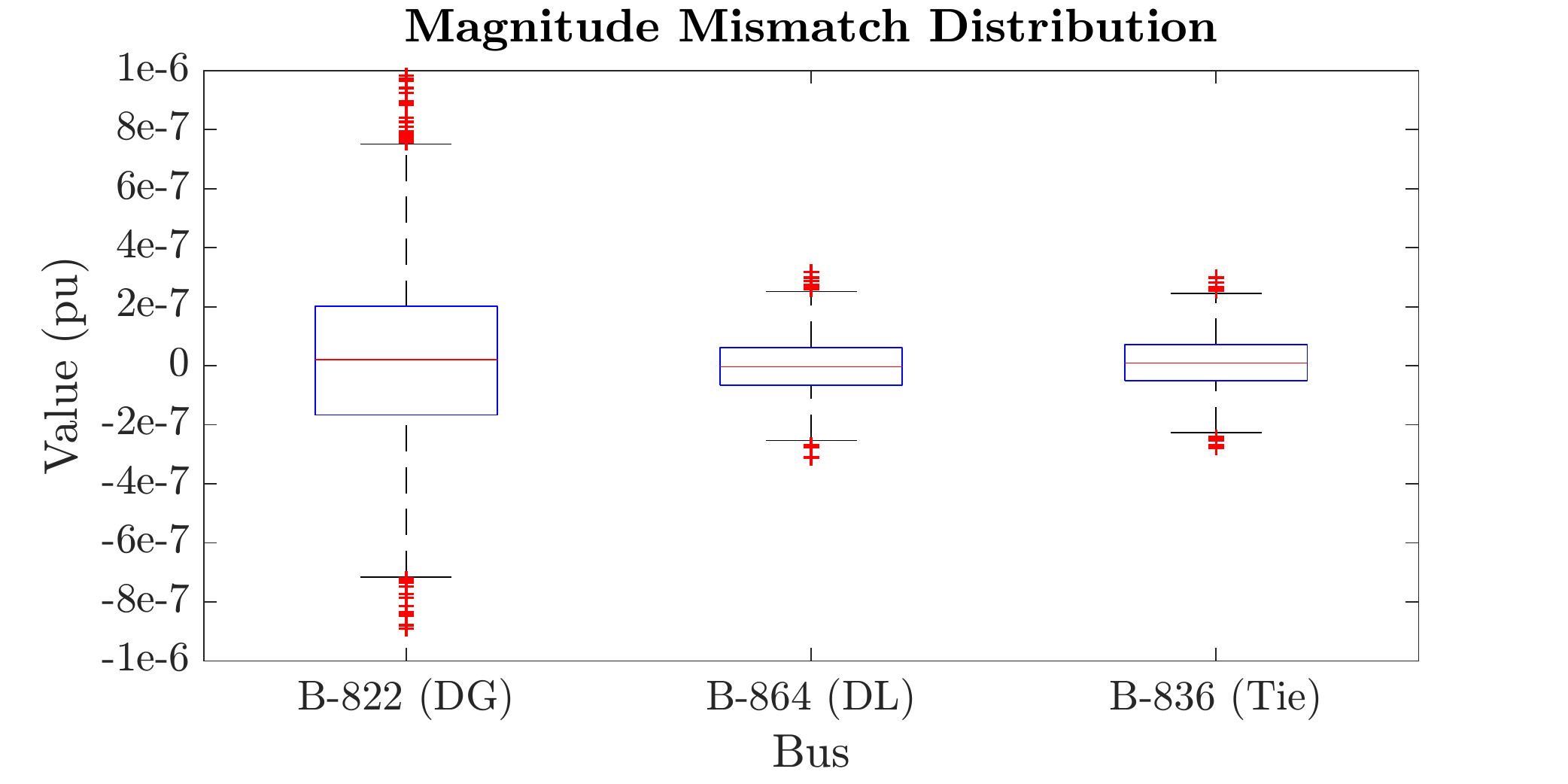}
		\label{Figure:Mismatch:Magnitude}
	}
	
	\subfloat[Phase mismatch $\Estimate{\delta}_{b,p,k}|_{\text{MUT}}-\Estimate{\delta}_{b,p,k}|_{\text{GM}}$.]
	{%
		\centering
		\includegraphics[width=1.0\linewidth]{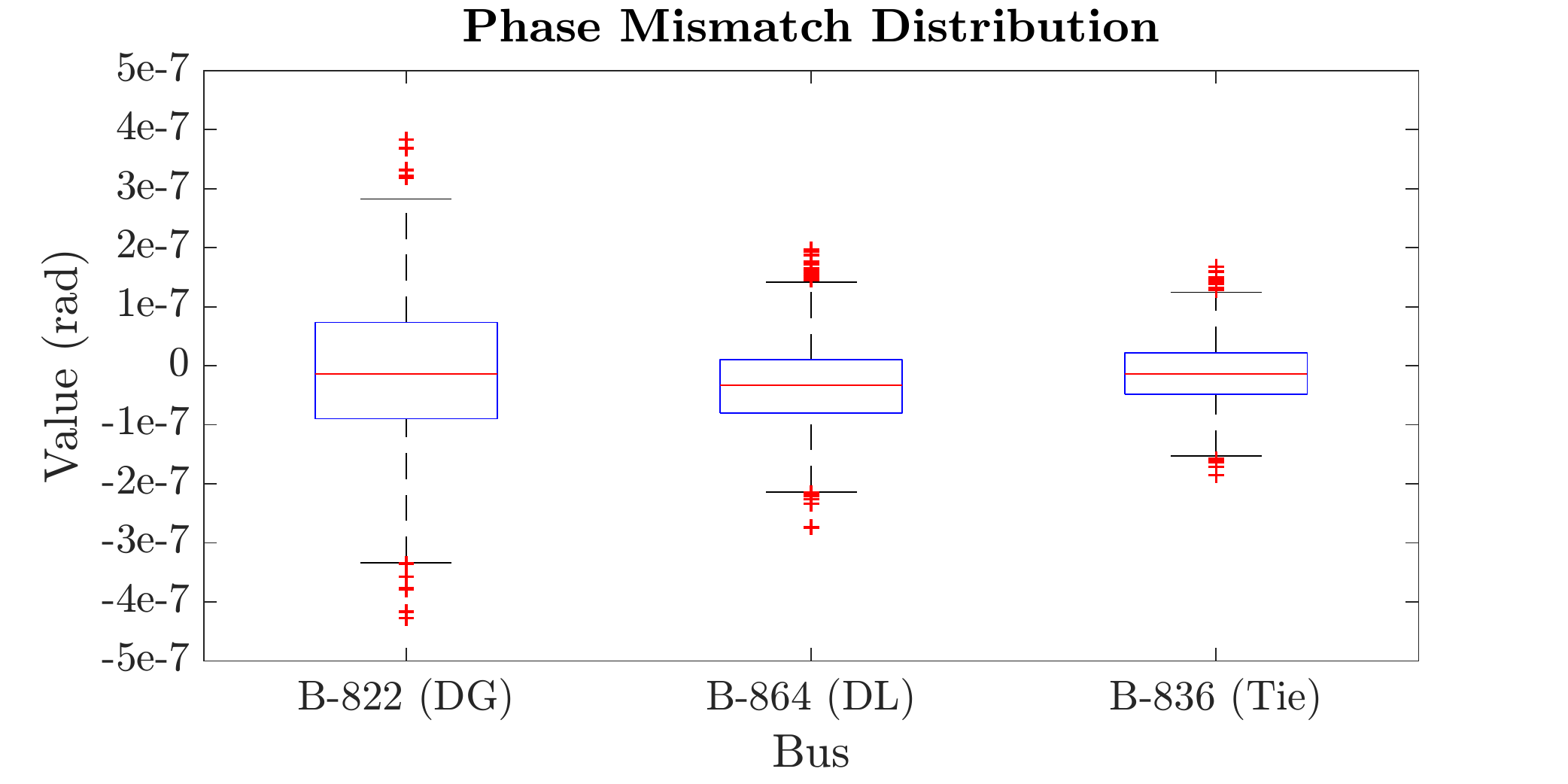}
		\label{Figure:Mismatch:Phase}
	}
	
	\caption{Distribution of the mismatch $\Estimate{V}_{b,p,k}|_{\text{MUT}}-\Estimate{V}_{b,p,k}|_{\text{GM}}$.}
	\label{Figure:Mismatch}
\end{figure}

It has been verified that the distribution of the error and mismatch quantities are static and close to normal,
which is in accordance with the assumptions made for the persistence process model \eqref{Equation:System:State:Model}
and the measurement model \eqref{Equation:System:Measurement:Model}.
The resulting distributions of the error and the mismatch are visualized in \Figure{Figure:Error}
and \Figure{Figure:Mismatch} for three different buses.
The sample data correspond to a time window of $40$ seconds (i.e. $2000$ samples at $50$ frames per second).
As one can see in \Figure{Figure:Error}, the estimation error is low both in magnitude and phase:
half of the samples are within $\pm2\cdot10^{-4}$ ($\PerUnit$ / $\Radian$).
This indicates that the \SDKF is tracking the state correctly,
and is in accordance with the performance assessment in \cite{Journal:Sarri:2016:SE:Linear:Performance}.
As \Figure{Figure:Mismatch} reveals, the results of the \MUT match well with those of the \GM.
Even at the bus with the largest mismatch, the magnitude and phase mismatch are within
$\pm1\cdot10^{-6}\PerUnit$ and $\pm5\cdot10^{-7}\Radian$, respectively.
Since the mismatch is substantially smaller than the modeled uncertainties, it can be concluded that the inaccuracy
due to the use of \SGL precision on the \FPGA, as compared to \DBL precision on the \CPU, is negligible.
In fact, since \SGL precision provides an accuracy of 6--7 decimal digits,
and a mismatch of $\leqslant 10^{-6}$ is expected.
Therefore, it can be concluded that the \MUT is equivalent to the \GM
within the bounds of the numerical accuracy.

\subsection{Scalability Analysis}
\label{Section:Validation:Scalability}


To assess the scalability of the \RTSE, the execution time of the \FPGA is measured
for estimation problems of different size.
%
Since benchmark feeders of arbitrary size are not readily available,
the necessary data are randomly generated, while ensuring that the working hypotheses of the \SDKF hold.
For simplicity, it is assumed that $D=S$, so an actual system would be observable with no redundancy%
\footnote{For this analysis, it is assumed that the matrix $\Tz_{k}$ is of full rank}.
The problem size is essentially limited by the amount of memory available on the \FPGA, namely $S<256$ if $D=S$.
Assuming an unreduced three-phase network, this corresponds to $N=255/(3\cdot2)\approx42$ nodes.
If network reduction techniques are used, for example the elimination of tie buses (applicable for any kind of network)
or the use of the single-phase equivalent (balanced networks only), considerably larger networks can be accommodated.
Note that execution time is defined as the time passing between the reading of the input and the writing of the output on the \FPGA.
In order to measure this time as accurately as possible, a counter driven by the master clock is implemented directly on the chip.
As the frequency of the master clock is known precisely, it is straightforward to derive the time from the counter state.


\begin{figure}[t]
	\centering
	\includegraphics[width=1.0\linewidth]{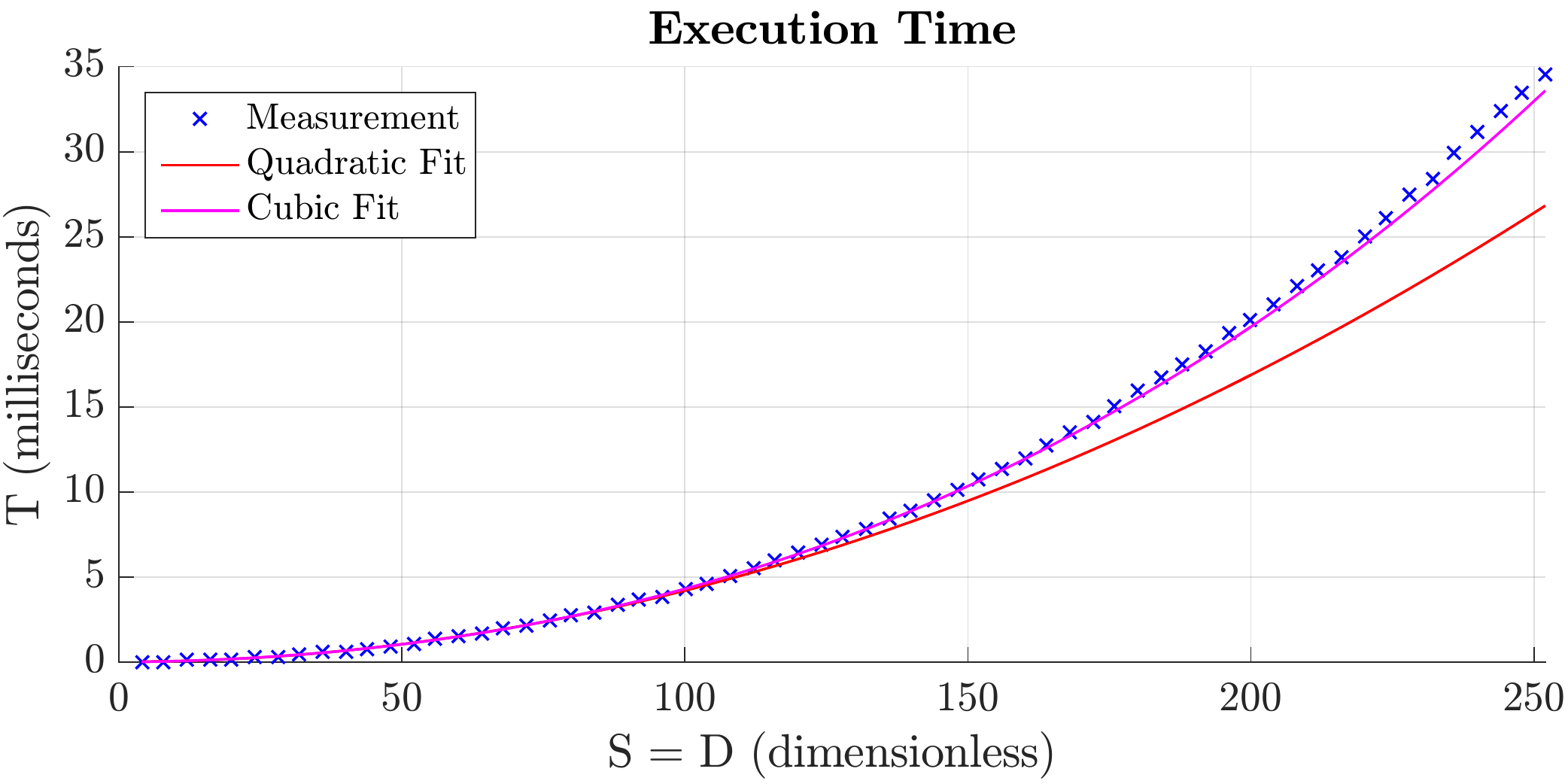}
	\caption{Execution time.}
	\label{Figure:Time}
\end{figure}

The obtained results are shown in \Figure{Figure:Time}.
As one can see, the time required for the largest problem size is 35\,ms.
In order to visualize the time complexity, a quadratic and a cubic curve are fit to the portion of the curve for which $S\leqslant80$.
As one would expect from the computational complexity analysis presented in \Section{Section:Algorithm:Complexity},
the time scales with the third power of the problem size.
However, it is worth noticing that the third order term is not dominant for this range of problem sizes,
since the cubic fit is not too far from the quadratic one.
This effect is due to the combination of parallelization and pipelining adopted for the implementation, 
as described in \Section{Section:Implementation:Prototype}.
For small problem sizes, the execution of the linear algebra blocks is dominated by the latency (the pipeline depth)
rather than the number of items to be processed.


\section{Conclusion}
\label{Section:Conclusion}

This paper has presented an \FPGA prototype of an \RTSE for \ADN[s] based on the \SDKF.
To motivate the use of the \SDKF rather than the \DKF, it has been proven that the two formulations are formally equivalent
(for uncorrelated measurement noise), and demonstrated that only the \SDKF is suitable for an implementation 
in this dedicated hardware.
To this effect, it has also been illustrated that the \SDKF only involves elementary linear algebra operations, 
which can be parallelized and pipelined in order to achieve high throughput.
The obtained results confirm that the developed \FPGA implementation of the \SDKF 
yields the same results as the reference \CPU implementation of the \DKF, while guaranteeing real-time performance.
In particular, the use of the \SGL number format on the \FPGA as opposed to the \DBL number format on the \CPU
does not cause any noteworthy inaccuracy.
Therefore, it can be concluded that \RTSE[s] on the basis of dedicated hardware implementations are indeed feasible,
and may hence support the development of automation systems for \ADN[s].


\appendices

\section{Transformation of the Uncertainty\\from Polar to Rectangular Coordinates}
\label{Appendix:Transformation}

Say $V=|V|\angle\delta$ the true value of a voltage phasor in polar coordinates.
Let $\Delta|V|$ and $\Delta\delta$ be the associated measurement errors,
so that the measured phasor $\widetilde{V}$ may be written as
\begin{equation}
	\widetilde{V} = (|V|+\Delta|V|)\angle(\delta+\Delta\delta)
\end{equation}
where $\Delta|V|$ and $\Delta\delta$ are assumed to be normally distributed
\begin{align}
	\Delta|V|		&\sim	\Normal{0,\sigma_{m}^{2}}\\
	\Delta\delta	&\sim	\Normal{0,\sigma_{p}^{2}}
\end{align}
According to \emph{Euler's Formula} $\widetilde{V}=\widetilde{V}_{r} + j\widetilde{V}_{i}$
\begin{align}
	\widetilde{V}_{r}	&= (|V|+\Delta|V|)\cos(\delta+\Delta\delta) = V_{r} + \Delta V_{r}\\
	\widetilde{V}_{i}	&= (|V|+\Delta|V|)\sin(\delta+\Delta\delta) = V_{i} + \Delta V_{i}
\end{align}
where $\Delta V_{r}$ and $\Delta V_{i}$ are the measurement errors in rectangular coordinates.
If $\Delta|V|$ and $\Delta\delta$ are independent, the variances $\sigma_{r}^{2}$ and $\sigma_{i}^{2}$
of $\Delta V_{r}$ and $\Delta V_{i}$ are given by \cite{Book:Milano:2016:Estimation}
\begin{align}
	\sigma_{r}^{2}
	&=	\left\{
			\begin{aligned}
				|V|^{2}e^{-\sigma_{p}^{2}}&\left[\cos^{2}\delta(\cosh^{2}(\sigma_{p}^{2})-1)+\sin^{2}\delta\sinh^{2}(\sigma_{p}^{2})\right]\\
				+\sigma_{m}^{2}e^{-\sigma_{p}^{2}}&\left[\cos^{2}\delta\cosh^{2}(\sigma_{p}^{2})+\sin^{2}\delta\sinh^{2}(\sigma_{p}^{2})\right]
			\end{aligned}
			\right.
			\label{Equation:Projection:Real}\\
	\sigma_{i}^{2}
	&=	\left\{
			\begin{aligned}
				|V|^{2}e^{-\sigma_{p}^{2}}&\left[\sin^{2}\delta(\cosh^{2}(\sigma_{p}^{2})-1)+\cos^{2}\delta\sinh^{2}(\sigma_{p}^{2})\right]\\
				+\sigma_{m}^{2}e^{-\sigma_{p}^{2}}&\left[\sin^{2}\delta\cosh^{2}(\sigma_{p}^{2})+\cos^{2}\delta\sinh^{2}(\sigma_{p}^{2})\right]
			\end{aligned}
			\right.
			\label{Equation:Projection:Imaginary}
\end{align}
Evidently, the uncertainties $\sigma_{r}$ and $\sigma_{i}$ in the rectangular coordinate system
do not only depend on the corresponding $\sigma_{m}$ and $\sigma_{p}$ in the polar coordinate system,
but also on the true magnitude and phase $|V|$ and $\delta$, which are unknown in practice.
Consider the case of the functional verification in \Section{Section:Validation:Functionality}.
A maximum measurement error of $1\cdot10^{-3}\PerUnit$ in magnitude
and $1.5\cdot10^{-3}\Radian$ in phase implies
\begin{align}
	\sigma_{m}	&\approx 3^{-1} \cdot 10^{-3}~\PerUnit\\
	\sigma_{p}		&\approx 5 \cdot 10^{-4}~\Radian
\end{align}
\Table{Table:Projection} lists $\sigma_{r}$ and $\sigma_{i}$ computed for $|V|=1~\PerUnit$
and different phase angles $\delta\in[0,\pi]~\Radian$.
According to \Equation{Equation:System:State:Vector}, \Equation{Equation:System:Measurement:Covariance},
and Hypothesis~\ref{Hypothesis:Uncorrelatedness}, $\sigma_{r}^{2}$ and $\sigma_{i}^{2}$
appear on the diagonal of $\Sz_{k}$.
If desired, $\Sz_{k}$ can thus be updated online based on the received measurements using the projection
defined by \Equation{Equation:Projection:Real} and \Equation{Equation:Projection:Imaginary}.
In this work, $\Sz_{k}=\Sz$ is presumed constant for the sake of simplicity (see \Section{Section:Validation:Functionality}).
For the transformation of $\sigma_{m}$ and $\sigma_{p}$ to $\sigma_{r}$ and $\sigma_{i}$,
it is assumed that the system is balanced, and that both the voltage drop and the phase angle difference
of any bus with respect to the slack are small.
That is, $\forall b\in\Buses$
\begin{equation}
	|V_{b,1}|=|V_{b,1}|=|V_{b,1}|\approx 1~\PerUnit
\end{equation}
\begin{equation}
	\delta_{b,1}\approx 0~\Radian,~
	\delta_{b,2}\approx -\frac{2\pi}{3}~\Radian,~
	\delta_{b,3}\approx +\frac{2\pi}{3}~\Radian
\end{equation}

\begin{table}
	\caption{Numerical example}
	\label{Table:Projection}
	\centering
	\renewcommand{\arraystretch}{1.5}
	\begin{tabular}{cccc}
		$|V|$ ($\PerUnit$)	&$\delta~(\Radian)$		&$\sigma_{r}$ ($\PerUnit$)	&$\sigma_{i}~(\PerUnit)$	\\
		\hline
		1							&$0$								&$3.333\cdot10^{-4}$			&$5.000\cdot10^{-4}$	\\
		1							&$\pm\frac{\pi}{6}$			&$3.819\cdot10^{-4}$			&$4.640\cdot10^{-4}$	\\
		1							&$\pm\frac{\pi}{3}$			&$4.640\cdot10^{-4}$			&$3.819\cdot10^{-4}$	\\
		1							&$\pm\frac{\pi}{2}$			&$5.000\cdot10^{-4}$			&$3.333\cdot10^{-4}$	\\
		1							&$\pm\frac{2\pi}{3}$		&$4.640\cdot10^{-4}$			&$3.819\cdot10^{-4}$	\\
		1							&$\pm\frac{5\pi}{6}$		&$3.819\cdot10^{-4}$			&$4.640\cdot10^{-4}$	\\
		1							&$\pi$							&$3.333\cdot10^{-4}$			&$5.000\cdot10^{-4}$
	\end{tabular}
\end{table}

\section{Strict Positive Definiteness\\of the Estimation Error Covariance}
\label{Appendix:Invertibility}

If the estimation error covariance matrix is initialized to be positive definite at start-up ($\Se_{0}^{+} \succ 0$),
$\Se_{k}^{-}$ and $\Se_{k}^{+}$ will remain positive definite for $k\geqslant1$, because this property is preserved
by the operations of the \DKF.
For the prediction step, this is straightforward to show.
From \Equation{Equation:Batch:Prediction:Covariance}, it is easy to see that
\begin{equation}
	\Se_{k-1}^{+}\succ0,~\Sx_{k}\succeq0\quad\Longrightarrow\quad \Se_{k}^{-}=\Se_{k-1}^{+}+\Sx_{k}\succ0
\end{equation}
For the estimation step, the following Lemma will be used.
\begin{Lemma}
	\label{Lemma:Definiteness}
	If $\Tensor{A}$ is a positive (semi)definite matrix, and $\Tensor{B}$ is an arbitrary matrix with full rank,
	then the matrix $\Tensor{C} = \Tensor{B}^{T}\Tensor{A}\Tensor{B}$ is also positive (semi)definite.
\end{Lemma}
Consider first the formulation \Equation{Equation:Batch:Estimation:Gain:A}--\Equation{Equation:Batch:Estimation:Covariance:A}
of the estimation step.
It should be noted that \Equation{Equation:Batch:Estimation:Covariance:A} is actually a simplified version
of a more complex symmetric expression, namely \cite{Book:Simon:2006:SE}
\begin{align}
	\Se_{k}^{+}
	&=	(\Tensor{I}-\K_{k}\Tz_{k})\Se_{k}^{-}	\\
	&=	(\Tensor{I}-\K_{k}\Tz_{k})\Se_{k}^{-}(\Tensor{I}-\K_{k}\Tz_{k})^{T} + \K_{k}\Sz_{k}\K_{k}^{T}
			\label{Equation:Definiteness:Symmetry}
\end{align}
Recall from \Equation{Equation:Batch:Estimation:Gain:A} that $\K_{k}$ is given by
\begin{equation}
	\K_{k} = \Se_{k}^{-}\Tz_{k}^{T}(\Tz_{k}\Se_{k}^{-}\Tz_{k}^{T}+\Sz_{k})^{-1}
\end{equation}
Since $\Tz_{k}$ has full rank (Hypothesis \ref{Hypothesis:Observability}), $\Sz_{k}$ is positive semidefinite by definition, 
and $\Se_{k}^{-}$ is positive definite by assumption, it follows that $(\ldots)^{-1}$ exists and that $\K_{k}$ has full rank.
By application of Lemma \ref{Lemma:Definiteness} to \Equation{Equation:Definiteness:Symmetry},
it follows that $\Se_{k}^{+}$ is positive definite.
Consider now the formulation \Equation{Equation:Batch:Estimation:Covariance:B}--\Equation{Equation:Batch:Estimation:State:B}
of the estimation step, which states that
\begin{equation}
	(\Se_{k}^{+})^{-1} = (\Se_{k}^{-})^{-1} + \Tz_{k}^{T}\Sz_{k}^{-1}\Tz_{k}
\end{equation}
If $(\Se_{k}^{-})$ is positive definite, its inverse exists and is positive definite as well.
The term $\Tz_{k}^{T}\Sz_{k}^{-1}\Tz_{k}$ is positive definite according to Lemma \ref{Lemma:Definiteness}.
By consequence, the sum term that defines $(\Se_{k}^{+})^{-1}$ is positive definite.

As the prediction step and the estimation step preserve the positive definiteness of $\Se_{k}^{-}$ and $\Se_{k}^{+}$
after an initialization with corresponding values, Hypothesis \ref{Hypothesis:Definiteness} is indeed reasonable.

\section{Discrete Kalman Filter Complexity}
\label{Appendix:Complexity}

\Table{Table:Batch:Detailed} and \Table{Table:Sequential:Detailed} list the number of operations required
for each step of the \DKF and the \SDKF, respectively.

{
\renewcommand{\arraystretch}{1.6}
\begin{table}[t]
	\centering
	\caption{Detailed Computational Complexity (\DKF)}
	\label{Table:Batch:Detailed}		
	\begin{tabular}{acc}
		\hline
		\multicolumn{2}{l}{\textbf{Prediction Step}}			&$+|-$							&$\times|\div$	\\
		\hline
		\xe_{k}^{-}		&= 	\xe_{k-1}^{+}							&$0$								&$0$	\\
		\Se_{k}^{-}		&=	\Se_{k-1}^{-} + \Sx_{k}			&$S$								&$0$	\\
		\hline
		\multicolumn{2}{l}{\textbf{Estimation Step}}			&$+|-$							&$\times|\div$	\\
		\hline
		\multicolumn{4}{l}{Reusable Coefficient}\\
		\dTz_{k}			&=	\Tz_{k}\Se_{k}^{-}					&$DS(S-1)$					&$DS^{2}$	\\
		\multicolumn{4}{l}{Kalman Gain}\\
		\dSz_{k}		&=	\dTz_{k}\Tz_{k}^{T}					&$D^{2}(S-1)$				&$D^{2}S$	\\
		\W_{k}			&=	\Sz_{k} + \dSz_{k}					&$D$							&$0$	\\
							&		\W_{k}^{-1}							&$m\in\Order{D^{3}}$		&$n\in\Order{D^{3}}$	\\
		\K_{k}			&= 	\dTz_{k}^{T}\W_{k}^{-1}			&$D(D-1)S$					&$D^{2}S$	\\
		\multicolumn{4}{l}{Estimated State}\\
		\ze_{k} 			&=	\Tz_{k}\xe_{k}^{-}					&$D(S-1)$						&$DS$	\\
		\dz_{k}			&= 	\za_{k} - \ze_{k}						&$D$							&$0$	\\
		\dx_{k}			&=	\K_{k}\dz_{k}							&$(D-1)S$						&$DS$	\\
		\xe_{k}^{+}		&=	\xe_{k}^{-} + \dx_{k}				&$S$								&$0$	\\
		\multicolumn{2}{l}{Estimation Error Covariance}	\\
		\dSx_{k}		&=	\K_{k}\dTz_{k}						&$(D-1)S^{2}$				&$DS^{2}$	\\
		\Se_{k}^{+}	&=	\Se_{k}^{-}-\dSx_{k}				&$S^{2}$						&$0$	\\
		\hline
	\end{tabular}
\end{table}
\begin{table}[t]
	\centering
	\caption{Detailed Computational Complexity (\SDKF)}
	\label{Table:Sequential:Detailed}		
	\begin{tabular}{acc}
		\hline
		\multicolumn{2}{l}{\textbf{Prediction Step}}			&$+|-$							&$\times|\div$	\\
		\hline
		\xe_{k}^{-}		&= 	\xe_{k-1}								&$0$								&$0$		\\
		\Se_{k}^{-}		&=	\Se_{k-1}^{+} + \Sx{k}				&$S$								&$0$		\\
		\hline
		\multicolumn{2}{l}{\textbf{Estimation Step}}			&$+|-$							&$\times|\div$	\\
		\hline
		\multicolumn{4}{l}{\textbf{FOR} $i\in\{1,\ldots,D\}$}\\
		\multicolumn{2}{l}{Reusable Coefficient}				&\phantom{$m\in\Order{D^{3}}$}	&\phantom{$n\in\Order{D^{3}}$}\\
		\dTz_{k,i}		&=	\Tz_{k,i}\Se_{k,i-1}^{+}			&$S(S-1)$						&$S^{2}$	\\
		\multicolumn{4}{l}{Kalman Gain}	\\
		\dSz_{k,i}		&=	\dTz_{k,i}\Tz_{k,i}^{T}				&$S-1$							&$S$		\\
		\W_{k,i}			&=	\Sz_{k,i} + \dSz_{k,i}				&$1$								&$0$		\\
							&		\W_{k,i}^{-1}							&$0$								&$1$		\\
		\K_{k,i} 			&= 	\dTz_{k,i}^{T}\W_{k,i}^{-1}		&$0$								&$S$		\\
		\multicolumn{4}{l}{Estimated State}	\\
		\ze_{k,i} 		&=	\Tz_{k,i}\xe_{k,i-1}^{+}				&$S-1$							&$S$	\\
		\dz_{k,i}			&= 	\za_{k,i} - \ze_{k,i}					&$1$								&$0$	\\
		\dx_{k,i}			&=	\K_{k,i}\dz_{k,i}						&$0$								&$S$	\\
		\xe_{k,i}^{+}	&=	\xe_{k,i-1}^{+} + \dx_{k,i}			&$S$								&$0$	\\
		%
		\multicolumn{2}{l}{Estimation Error Covariance}		\\
		\dSx_{k,i}		&=	\K_{k,i}\dTz_{k,i}						&$0$								&$S^{2}$	\\
		\Se_{k,i}^{+}	&=	\Se_{k,i-1}^{+}-\dSx_{k}			&$S^{2}$						&$0$	\\
		\multicolumn{4}{l}{\textbf{END}}	\\
		\hline
	\end{tabular}
\end{table}
}

\section*{Acknowledgment}

This work has been funded by the National Research Programme NRP70 ``Energy Turnaround''
of the Swiss National Science Foundation (SNSF).
For further information, please refer to \emph{www.nrp70.ch}.	


\ifCLASSOPTIONcaptionsoff
	\newpage
\fi



%


\bibliographystyle{IEEEtran} 
\bibliography{Bibliography}


\begin{IEEEbiography}[{\includegraphics[width=2.6cm,keepaspectratio]{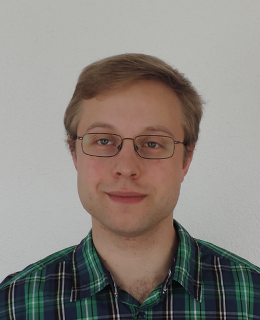}}]{Andreas Martin Kettner}
	(M'15) grew up in S{\"u}nikon, Switzerland, 
	and attended the Kantonsschule Z{\"u}rcher Unterland in B{\"u}lach, Switzerland.
	He received the B.Sc. and M.Sc. degrees in Electrical Engineering and Information Technology from 
	the Eidgen{\"o}ssische Technische Hochschule Z{\"u}rich, Switzerland,
	in 2012 and 2014, respectively.
	After working as a development engineer for Supercomputing Systems AG in Z{\"u}rich,
	he joined the Distributed Electrical Systems Laboratory at the
	{\'E}cole Polytechnique F{\'e}d{\'e}rale de Lausanne, Switzerland,
	where he is pursuing a Ph.D. degree.
	His research interests include real-time monitoring and control of Active Distribution Networks
	with particular focus on State Estimation and Voltage Stability Assessment.
\end{IEEEbiography}

%

\begin{IEEEbiography}[{\includegraphics[width=2.6cm,keepaspectratio]{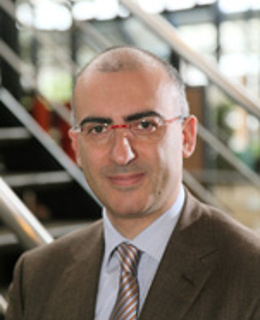}}]{Mario Paolone}
	(M'07--SM'10) received the M.Sc. (Hons.) and Ph.D. degrees in electrical engineering
	from the University of Bologna, Bologna, Italy, in 1998 and 2002, respectively.
	In 2005, he was nominated Assistant Professor in power systems at the University of Bologna,
	where he was with the power systems laboratory until 2011.
	He is currently an Associate Professor at the Swiss Federal Institute of Technology, Lausanne, Switzerland,
	chair of the Distributed Electrical Systems Laboratory.
	He was the Co-Chairperson of the Technical Committee of the 9th edition of the International Conference of Power Systems Transients (2009)
	and of the 19th and 20th Power Systems Computation Conference (2016 and 2018).
	He is author or coauthor of more than 220 scientific papers published in reviewed journals and international conferences.
	He is the Editor-in-Chief of the Elsevier journal Sustainable Energy, Grids and Networks
	and the Head of the Swiss Competence Center for Energy Research ``FURIES''.
	His research interests include power systems with particular reference to real-time monitoring and operation of active distribution networks,
	integration of distributed energy storage systems, power system protections and power system transients.
	In 2013, he received the IEEE EMC Society Technical Achievement Award. 
\end{IEEEbiography}

%
%
%
%
%
%
%
%

\vfill


\end{document}